\documentclass[letterpaper,twocolumn,prl,aps,superscriptaddress,amsmath,amssymb,floatfix]{revtex4-1}
\usepackage{mathptmx}
\usepackage[utf8]{inputenc} 
\usepackage[T1]{fontenc}    
\usepackage{color}
\usepackage{amsmath}
\usepackage{amssymb}
\usepackage{graphicx}
\usepackage{esint}
\usepackage[unicode=true,
bookmarks=true,bookmarksnumbered=false,bookmarksopen=false,
breaklinks=false,pdfborder={0 0 1},backref=false,colorlinks=true]
{hyperref}
\hypersetup{
	linkcolor=magenta,urlcolor=blue,citecolor=blue,pdfstartview={FitH},hyperfootnotes=false}

\makeatletter

\usepackage{textcomp}
\usepackage{epstopdf}

\pdfpageheight\paperheight
\pdfpagewidth\paperwidth

\@ifundefined{textcolor}{}{%
	\definecolor{BLACK}{gray}{0}
	\definecolor{WHITE}{gray}{1}
	\definecolor{RED}{rgb}{1,0,0}
	\definecolor{GREEN}{rgb}{0,1,0}
	\definecolor{BLUE}{rgb}{0,0,1}
	\definecolor{CYAN}{cmyk}{1,0,0,0}
	\definecolor{MAGENTA}{cmyk}{0,1,0,0}
	\definecolor{YELLOW}{cmyk}{0,0,1,0}
}

\usepackage{xcolor}\usepackage{soul}
\definecolor{blue}{rgb}{0,0,1}
\definecolor{red}{rgb}{1,0,0}
\definecolor{green}{rgb}{0,1,0}

\usepackage{siunitx}
\usepackage{mathrsfs}

\makeatother

\begin{document}
\title{Reconfigurable on-chip vortex beam generation via acoustically stimulated Brillouin nonlinear optical radiation}
\author{Ming Li}
\thanks{These authors contribute equally to this work.}
\affiliation{Laboratory of Quantum Information, 
University of Science and Technology of China, Hefei 230026, China}
\affiliation{CAS Center For Excellence in Quantum Information and Quantum Physics,
University of Science and Technology of China, Hefei, Anhui 230026,
China}
\affiliation{Anhui Province Key Laboratory of Quantum Network, 
University of Science and Technology of China, Hefei 230026, China}
\affiliation{Hefei National Laboratory, Hefei 230088, China}

\author{Xiang Chen}
\thanks{These authors contribute equally to this work.}
\affiliation{Laboratory of Quantum Information, 
University of Science and Technology of China, Hefei 230026, China}
\affiliation{CAS Center For Excellence in Quantum Information and Quantum Physics,
University of Science and Technology of China, Hefei, Anhui 230026,
China}
\affiliation{Anhui Province Key Laboratory of Quantum Network, 
University of Science and Technology of China, Hefei 230026, China}

\author{Wen-Qi Duan}
\affiliation{Laboratory of Quantum Information, 
University of Science and Technology of China, Hefei 230026, China}
\affiliation{CAS Center For Excellence in Quantum Information and Quantum Physics,
University of Science and Technology of China, Hefei, Anhui 230026,
China}
\affiliation{Anhui Province Key Laboratory of Quantum Network, 
University of Science and Technology of China, Hefei 230026, China}

\author{Yuan-Hao Yang}
\affiliation{Laboratory of Quantum Information, 
University of Science and Technology of China, Hefei 230026, China}
\affiliation{CAS Center For Excellence in Quantum Information and Quantum Physics,
University of Science and Technology of China, Hefei, Anhui 230026,
China}
\affiliation{Anhui Province Key Laboratory of Quantum Network, 
University of Science and Technology of China, Hefei 230026, China}

\author{Jia-Qi Wang}
\affiliation{Laboratory of Quantum Information, 
University of Science and Technology of China, Hefei 230026, China}
\affiliation{CAS Center For Excellence in Quantum Information and Quantum Physics,
University of Science and Technology of China, Hefei, Anhui 230026,
China}
\affiliation{Anhui Province Key Laboratory of Quantum Network, 
University of Science and Technology of China, Hefei 230026, China}

\author{Mai Zhang}
\affiliation{Laboratory of Quantum Information, 
University of Science and Technology of China, Hefei 230026, China}
\affiliation{CAS Center For Excellence in Quantum Information and Quantum Physics,
University of Science and Technology of China, Hefei, Anhui 230026,
China}
\affiliation{Anhui Province Key Laboratory of Quantum Network, 
University of Science and Technology of China, Hefei 230026, China}

\author{Xin-Biao Xu}
\affiliation{Laboratory of Quantum Information, 
University of Science and Technology of China, Hefei 230026, China}
\affiliation{CAS Center For Excellence in Quantum Information and Quantum Physics,
University of Science and Technology of China, Hefei, Anhui 230026,
China}
\affiliation{Anhui Province Key Laboratory of Quantum Network, 
University of Science and Technology of China, Hefei 230026, China}

\author{Chun-Hua Dong}
\affiliation{Laboratory of Quantum Information, 
University of Science and Technology of China, Hefei 230026, China}
\affiliation{CAS Center For Excellence in Quantum Information and Quantum Physics,
University of Science and Technology of China, Hefei, Anhui 230026,
China}
\affiliation{Anhui Province Key Laboratory of Quantum Network, 
University of Science and Technology of China, Hefei 230026, China}
\affiliation{Hefei National Laboratory, Hefei 230088, China}

\author{Luyan Sun}
\affiliation{Center for Quantum Information, Institute for Interdisciplinary Information Sciences, Tsinghua University, Beijing 100084, China}
\affiliation{Hefei National Laboratory, Hefei 230088, China}

\author{Guang-Can Guo}
\affiliation{Laboratory of Quantum Information, 
University of Science and Technology of China, Hefei 230026, China}
\affiliation{CAS Center For Excellence in Quantum Information and Quantum Physics,
University of Science and Technology of China, Hefei, Anhui 230026,
China}
\affiliation{Anhui Province Key Laboratory of Quantum Network, 
University of Science and Technology of China, Hefei 230026, China}
\affiliation{Hefei National Laboratory, Hefei 230088, China}

\author{Chang-Ling Zou}
\email{clzou321@ustc.edu.cn}
\affiliation{Laboratory of Quantum Information, 
University of Science and Technology of China, Hefei 230026, China}
\affiliation{CAS Center For Excellence in Quantum Information and Quantum Physics,
University of Science and Technology of China, Hefei, Anhui 230026,
China}
\affiliation{Anhui Province Key Laboratory of Quantum Network, 
University of Science and Technology of China, Hefei 230026, China}
\affiliation{Hefei National Laboratory, Hefei 230088, China}

\date{\today}
\begin{abstract}
The integrated devices that generate structured optical fields with non-trivial orbital angular momentum (OAM) hold great potential for advanced optical applications, but are restricted to complex nanostructures and static functionalities. Here, we demonstrate a reconfigurable OAM beam generator from a simple microring resonator without requiring grating-like nanostructures. Our approach harnesses Brillouin interaction between confined phonon and optical modes, where the acoustic field is excited through microwave input. The phonon stimulate the conversion from a guided optical mode into a free-space vortex beam. Under the selection rule of the radiation, the OAM order of the emitted light is determined by the acousto-optic phase matching and is rapidly reconfigurable by simply tuning the microwave frequency. Furthermore, this all-microwave control scheme allows for the synthesis of arbitrary high-dimensional OAM superpositions by programming the amplitudes and phases of the driving fields. Analytical and numerical models predict a radiation efficiency over 25\% for experimentally feasible on-chip microcavities. This work introduces a novel paradigm for chip-to-free-space interfaces, replacing fixed nanophotonic structures with programmable acousto-optic interactions for versatile structured light generation.
\end{abstract}
\maketitle

\noindent
\textit{Introduction.--}
Vortex optical beams carrying OAM represent a powerful tool with diverse applications, ranging from high-capacity
classical and quantum communication~\cite{wang2012terabit,molina2007twisted,fickler2012quantum,fickler2016quantum,stav2018quantum,erhard2020advances,ding2015quantum,mirhosseini2015high} to optical manipulation~\cite{grier2003revolution,padgett2011tweezers,paterson2001controlled,shen2019optical} and optical microscopy~\cite{furhapter2005spiral,PhysRevLett.97.163903}. While conventional optical components, such as digital micromirror devices,
holograms, Q-plates, and liquid-crystal spatial light modulators~\cite{yao2011orbital, forbes2021structured,chen2021engineering}, have been widely employed to generate these beams, integrated photonics devices have attracted great attention. The photonic chip offers a promising path toward scalable and robust vortex beam devices~\cite{yang2021perspective}.
Engineered nanostructures, such as
metasurfaces~\cite{scienceOAMspin,sroor2020high,Yang:21}, plasmonic structures~\cite{C4NR03606A}, and gratings~\cite{liu2016chip,caiscience,zhao2025all} have been utilized to transform on-chip waveguide modes or resonant whispering gallery modes (WGMs)~\cite{strekalov2016nonlinear} into free-space vortex beams. Significant advancements have been achieved, including the demonstration of vortex lasers~\cite{miao2016orbital},
entangled vortex photon pairs~\cite{chen2021bright}, and vortex microcombs~\cite{liu2024integrated, chen2024integrated}. 

Despite these advancements, existing on-chip OAM beam emitters face critical limitations that hinder their practical applications. First, they rely on distributed scatterers with certain patterns for scattering the input optical field, featuring fixed, complex nanostructures that are sensitive to fabrication imperfections and lack the capability for rapid reconfiguration. Second, generating and controlling arbitrary high-dimensional OAM superposition states, which is a key resource for advanced information technologies~\cite{fickler2012quantum,fickler2016quantum,erhard2020advances}, remains a significant challenge. Beyond emitters dependent on linear scattering processes, recent studies have demonstrated that nonlinear optics effects, such as parametric $\chi^{(2)}$ nonlinearity and Brillouin scattering, can induce the direct radiation of on-chip guided photons into free space by a uniform photonic structure~\cite{NOR2023,li2023frequency}. These nonlinear optical radiation (NOR) processes are controlled by ancillary optical or acoustic fields, thus providing opportunities for flexible reconfiguration of the radiation fields. 

\begin{figure}
\centering
\includegraphics[width=1\columnwidth]{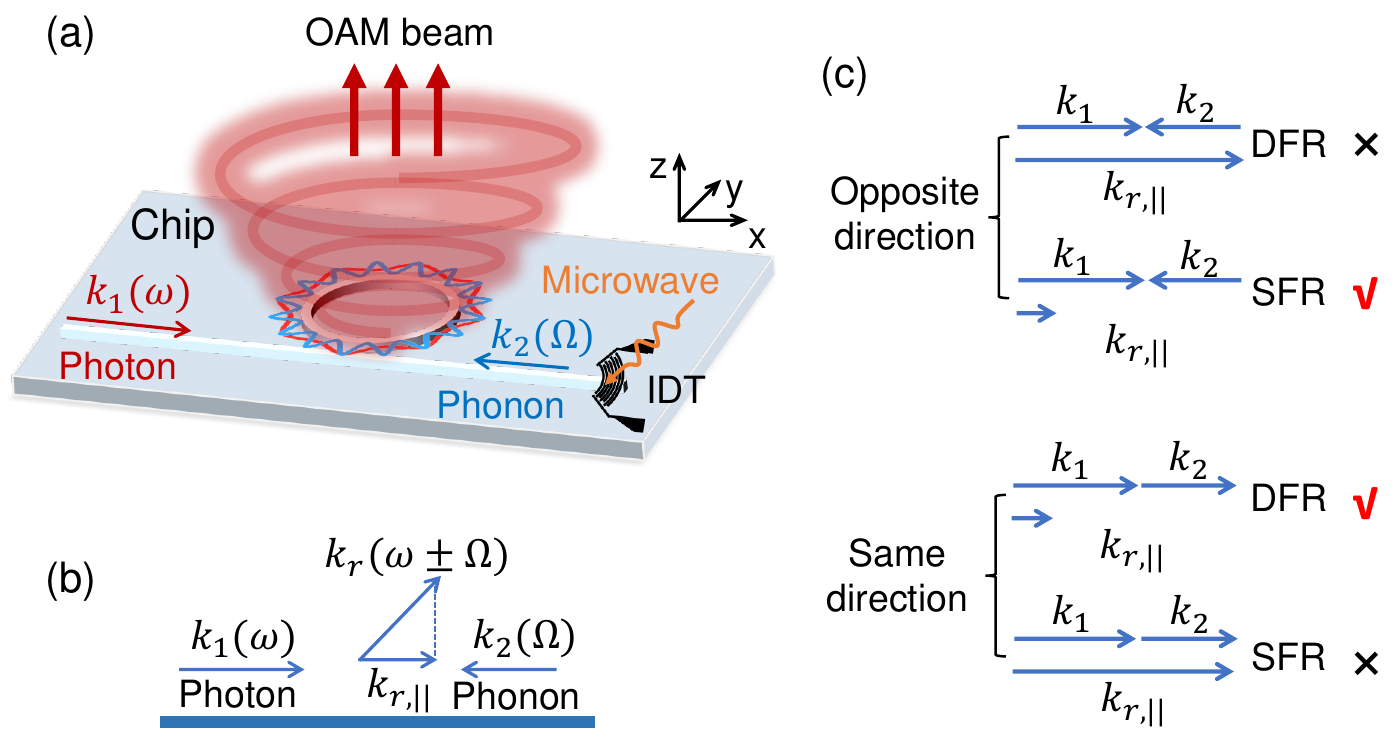}
\caption{Principle of Brillouin NOR and OAM beam radiation. (a) Radiation of OAM beams from a microring cavity supporting both photonic and phononic resonant modes. The phonons are excited by microwave using IDT through piezoelectric effect. $k_{1}(\omega)$ and $k_{2}(\Omega)$ are the wave vectors of the photonic and phononic modes in the waveguide.  IDT: interdigital
transducer. (b) Illustration of the wave vectors of the guided photons, guided phonons and free-space photons along the waveguide. (c) Selection rules of the radiation. Four types of radiation processes between one photonic
mode and one phononic mode. SFR: sum-frequency radiation. DFR: difference-frequency
radiation.}
\label{Fig1}
\end{figure}

In this Letter, we introduce a fundamentally new mechanism for OAM beam generation based on Brillouin NOR in a grating-free microring resonator. In our system, a microwave-driven phononic mode directly stimulates the coupling between the confined optical field to free-space radiation, thereby imprinting the net OAM of the resonant modes onto the emitted beam. This acousto-optic interaction not only eliminates the need for any fixed nanostructures but also provides complete, real-time control over the output. The OAM order can be rapidly tuned by sweeping the microwave frequency, and arbitrary, high-dimensional OAM superpositions can be synthesized by simply programming the amplitudes and phases of the driving microwave fields. Our theoretical model, validated by numerical simulations, predicts a radiation efficiency exceeding 25\% in experimentally feasible microcavities~\cite{Liu2022,Xu2025a} supporting confined phonon modes for Brillouin scattering~\cite{Yang2024ome,Ye2025,Rodrigues2025,Yu2025,Yang2025a,Xu2025b,Chen2025}.
This work establishes a new interface for chip-to-free-space optical conversions, and paves the way for dynamically reconfigurable and programmable generation, manipulation, and detection of vortex beams by integrated photonic devices, thereby advancing high-capacity optical interconnects and quantum information processing.

\noindent
\textit{Brillouin NOR for OAM radiation.--}
Figure~\ref{Fig1}(a) illustrates the core idea of our proposal: reconfigurable structured light generation through dynamic acousto-optic interactions in a simple on-chip microring resonator that simultaneously supports both photonic and phononic WGMs~\cite{Yang2023,Yang2025a}, with optical band photons and microwave band phonons circulating along the waveguides. The device harnesses backward Brillouin interaction that induces a nonlinear polarization $P_{NL}\propto d_{j}(r) P s_{n}(r)$, where $P$ is the photoelastic coefficient, and {$d_j$ and $S_n$ are the optical and acoustic displacement fields, respectively}. This nonlinear polarization acts as a distributed radiation source, similar to the previously demonstrated all-optical NOR based on sum-frequency generation in Ref.~\cite{NOR2023}, driving photons from the guided optical mode into free-space radiation [Fig.~\ref{Fig1}(a)]. Brillouin NOR has a crucial distinction from the $\chi^{(2)}$-based NOR~\cite{NOR2023} in that it involves the mixing of optical and acoustic fields. Since the acoustic modes have much slower velocity compared to the optical modes, the optical modes can simultaneously couple with multiple different acoustic modes, and two fundamental acousto-optic processes are possible: \textit{sum-frequency radiation} (SFR), where a guided photon absorbs a phonon to radiate a higher-frequency photon into free space ($\omega_{\mathrm{out}} = \omega + \Omega$), and \textit{difference-frequency radiation} (DFR), where photons stimulate phonon emission and radiates a lower-frequency photon ($\omega_{\mathrm{out}} = \omega - \Omega$).

Not all combinations of optical and acoustic modes lead to efficient radiation, and the key to efficient NOR is momentum conservation. In particular, considering a section of straight waveguide, the in-plane coupling between propagating optical and acoustic waves requires momentum conservation parallel to the waveguide, while the requirement for the vertical direction is negligible due to the very thin interaction region. As illustrated in Fig.~\ref{Fig1}(b), the phase-matching condition reads $k_{1}\pm k_{2}=k_{r,||}$, where $k_{1}$, $k_{2}$, and $k_{r,||}$ are the in-plane wave vector components of the guided photon, phonon and radiated photon, respectively. For propagating waves, the radiated photon must obey $|k_{r,||}|\leq k_0$ while $k_{1}\approx n_{\mathrm{eff}}k_0$, where $k_0$ is the wave number in vacuum and $n_{\mathrm{eff}}$ is the mode's effective refractive index, which should be larger than the refractive index of the substrate material. Then, a stringent selection rule emerges: two of the four possible interaction configurations shown in Fig.~\ref{Fig1}(c) are forbidden, i.e., the DFR of counter-propagating modes and the SFR of co-propagating modes, as they would require $|k_{r,||}|> k_r$. Consequently, only two processes lead to efficient radiation: SFR of counter-propagating modes and DFR of co-propagating modes. This imposes a requirement on the phonon wave vector:
\begin{equation}
    \left(1 - \frac1{n_\mathrm{eff}}\right) k_1 \lesssim k_2 \lesssim \left(1 + \frac1{n_\mathrm{eff}}\right) k_1. 
    \label{eq:wavevector}
\end{equation}
This condition is critical, as it defines the necessary microwave frequency range, via the phononic dispersion $k_{2}(\Omega)$ of the waveguide, to activate the Brillouin NOR.

Building on the basic principle of Brillouin NOR, the radiated field distributed under an acoustic drive field can be derived and numerically calculated (see \cite{sm} for details). Here, the phononic modes are excited by applying a microwave signal to interdigital transducers (IDTs), which leverages the material's piezoelectricity to drive a specific acoustic resonance. According to the resonator's rotational symmetry, the far-field radiation pattern can be calculated from~\cite{sm} 
\begin{eqnarray}
E_{j,n}(r) \propto \alpha_{1}\alpha_{2} e^{i(m_{j}\pm M_{n})\phi},
\label{eq:radiationfield}
\end{eqnarray}
where $\alpha_{1,2}$ are the in-plane photon and phonon mode amplitudes, $m_{j}$ and $M_n$ label the azimuthal OAM of the modes, which relate to the cavity radius $R$ by $m_{j} \approx k_{1,j} R$ for photons and $M_{j} \approx k_{2,j} R$ for phonons, and $\phi$ is the azimuth angle. This 
equation shows that the resulting radiation is a vortex beam with a well-defined OAM order of $m_{j}\pm M_{n}$, with the sign determined by the radiation process (SFR or DFR). Therefore, by selecting the initial photonic mode ($m_j$) with a laser and the phononic mode ($M_n$) with a microwave signal, our approach provides a unique approach to realize electrically-stimulated NOR for any desired OAM order with a wide range of tunable frequencies.

\begin{figure*}
\begin{centering}
\includegraphics[width=\textwidth]{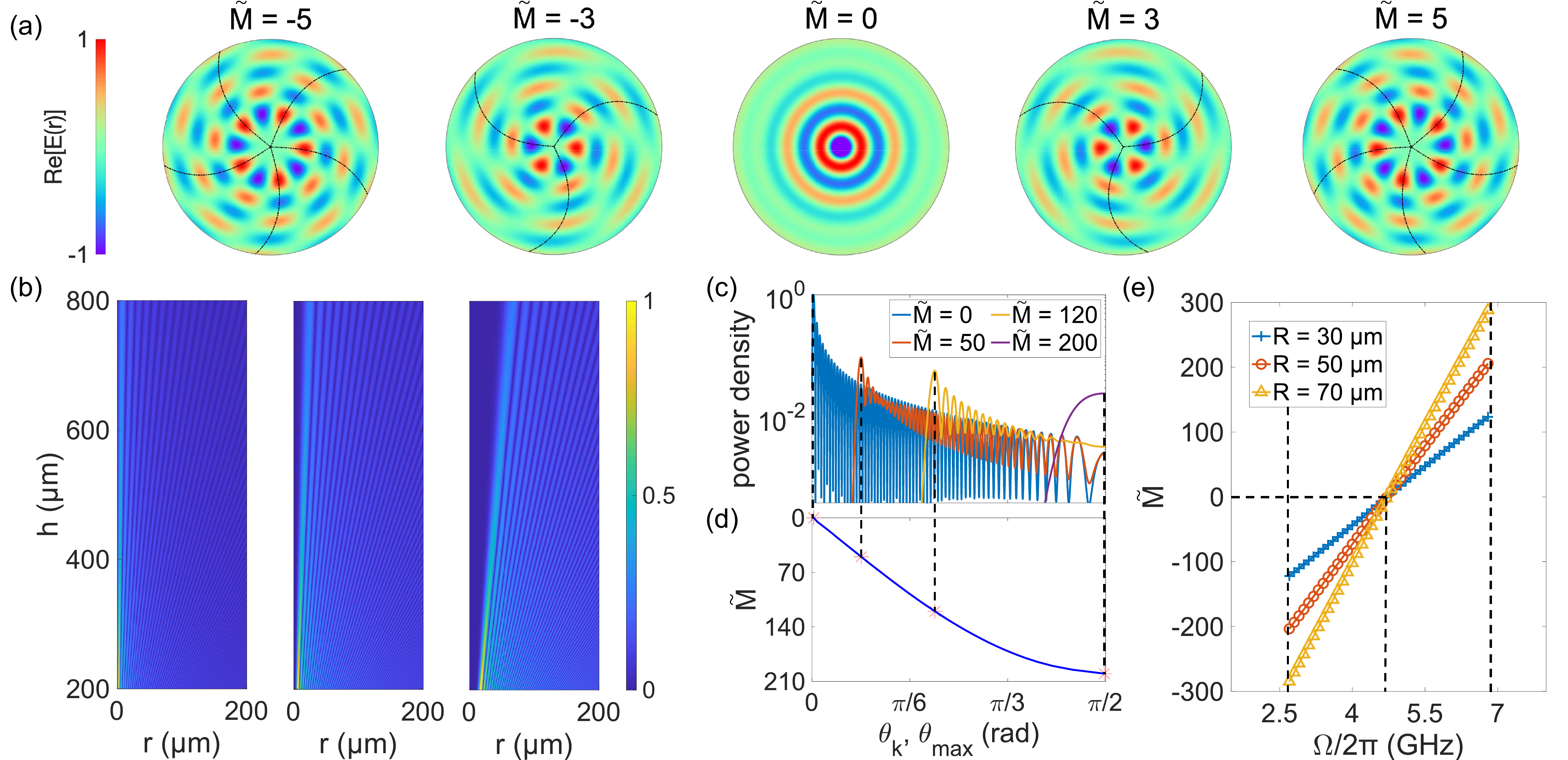}
\par\end{centering}
\caption{Field distribution of OAM beam radiated from a microring cavity. 
(a) Real part of the radiation field at $z = \SI{100}{\micro m}$ in the
XY plane with a radius of $\SI{10}{\micro m}$ for OAM orders $-5,-3,0,3,5$, respectively. 
(b) The power density distribution in the $r$-space. $r$ is the distance from the ring center. (c) The power density distribution in the $k$-space. $\theta$ is the angle of the wave vector $k$ to z-axis.
(d) Relationship between the angle $\theta_{\mathrm{max}}$ with maximal intensity and OAM order $\widetilde M$.
(e) Relationship between the OAM order $\widetilde M$ and the phonon frequency for $R =$ 30, 50, and 70$\,\SI{}{\micro m}$. The dashed lines show the frequency range of the phonon mode restricted by Eq.~(\ref{eq:wavevector}), which bounds the maximum OAM order for different ring radius $R$. 
}
\label{Fig2}
\end{figure*}

\smallskip{}
\noindent
\textit{OAM generation.--}
Figure~\ref{Fig2} presents numerical results that validate our theoretical model (see Supplemental Materials for geometric parameters~\cite{sm}). We focus on the SFR involving a clockwise (CW) photonic mode of order $m$ and counter-clockwise (CCW) phononic mode of order $-M$. The calculated far-field patterns of $\text{Re}(E)$ at $100\,\mathrm{\mu m}$ above the microring with a radius of $\SI{10}{\micro m}$, shown in Fig.~\ref{Fig2}(a) for $\widetilde M = 0, \pm3, \pm5$, clearly exhibit the characteristic spiral phase fronts of vortex beams. The number of azimuthal phase cycles matches the OAM order $\widetilde M=m-M$ exactly as expected, with the spiral matching the sign of $\widetilde M$, confirming the successful transfer of OAM from on-chip modes to the radiated field. The corresponding field distributions in both real space [Fig.~\ref{Fig2}(b)] and momentum space [Fig.~\ref{Fig2}(c)] display the distinctive annular (hollow) profile characteristic of vortex beams. The momentum is characterized by the angle $\theta_{k}$ with respect to the vertical direction ($z$-axis). The radius of the central dark core expands with increasing OAM order $|\widetilde M|$, as shown in Fig.~\ref{Fig2}(b) and marked by the dashed lines in Fig.~\ref{Fig2}(c). This expansion of the vortex beam is directly linked to the momentum conservation principle, since the out-of-plane momentum reduces with increasing OAM, thus the angle of maximum radiation intensity ($\theta_{\text{max}}$) reduces, as shown in Fig.~\ref{Fig2}(d).

The phase-matching condition in Eq.~(\ref{eq:wavevector}) also sets the operational frequency range and an upper limit on the achievable OAM order. Figure~\ref{Fig2}(e) shows the relationship between the generated OAM order $\widetilde M=m-M$ and the required phonon frequency for resonators of different radii. For low-order OAM generation, the required microwave frequencies are around half of the Brillouin scattering frequencies in materials, according to the backward Brillouin scattering ($4.5\,\mathrm{GHz}$ for $1550\,\mathrm{nm}$ light in lithium niobate waveguide~\cite{Yang2023}). 
For a microring resonator of radius $R$, the generated vortex beam should satisfy the relation $|\widetilde{M}|\leq k_0 R$ for vertical propagation, thus the maximally achievable OAM order $|\widetilde M|_\mathrm{max} \approx k_0 R$, which also determines the required phonon momentum $|k_2-k_1|\leq k_0$, leading to a fixed range of phonon frequency according to the phononic dispersion in the waveguide, as shown by the dashed lines in Fig.~\ref{Fig2}(e).

\smallskip{}
\noindent
\textit{Radiation efficiency.--}
The radiation efficiency is the critical metric for OAM beam generators. In conventional on-chip devices, the radiation efficiency is a fixed value that determined by the fabricated nanostructures. In contrast, the efficiency of our grating-free Brillouin NOR scheme is a dynamic quantity that depends on the driving phonon mode amplitude. The radiation efficiency $\eta$ is defined as the ratio of the radiation optical power to the input optical power. Since the driving phonons stimulate the coupling between optical cavity mode and the free-space radiation, the efficiency can be calculated through ~\cite{sm}
\begin{equation} 
    \eta =  \frac{4 \kappa_{\text{nl}} \kappa_{\text{ex}}}{(\kappa_{\text{nl}} + \kappa_{\text{ex}} + \kappa_\text{int})^2}, 
    \label{equ_P_kappa}
\end{equation}
where  $\kappa_{\text{nl}} = \kappa_{2} |\alpha_2|^2$ is the phonon-number-dependent nonlinear decay rate due to Brillouin NOR, with $\kappa_2=\frac{P_{\mathrm{rad}}/\hbar \omega }{2|\alpha_1\alpha_2|^2}$ being the intensity-normalized nonlinear coupling rate that only depends cavity geometry (see Ref.~\cite{sm} for detailed derivations). $\kappa_\text{ex}$ and $\kappa_\text{int}$ are the standard linear external coupling and intrinsic loss rates of the photon mode. We note that the efficiency is critically dependent on the photonic ($Q_1$) and phononic ($Q_2$) quality factors, as $\kappa_\mathrm{int}+\kappa_\mathrm{ex}\propto1/Q_1$ and intracavity phonon amplitude $ |\alpha_2|^2\propto 1/Q_2$ could be drastically enhanced by the phonon mode. Additionally, $\kappa_2$ inversely scales with the phonon mode volume $\kappa_{2}\propto R^{-1}$, since the light-matter interaction can be enhanced by tightly confined modes in microresonators~\cite{PhysRevX.14.011056,Liu2022}. 

\begin{figure}
\centering
\includegraphics[width=1\columnwidth]{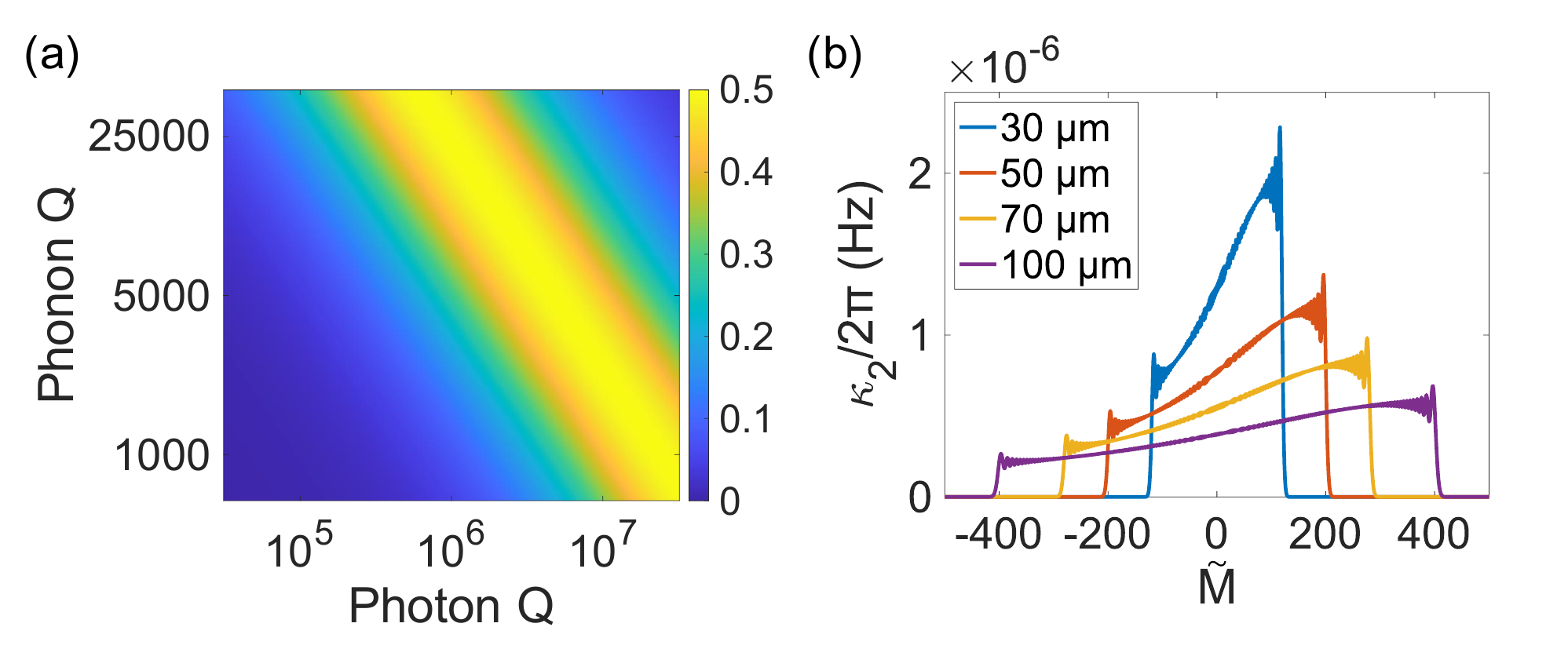}
\caption{Radiation efficiency of OAM beam. (a) Relationship between the radiation efficiency and the quality factors of resonant modes in a $50\,\mathrm{\mu m}$-radius microring. The ring is fixed at critical coupling for both photon and phonon modes. The input powers are $P_{\mathrm{phonon}} = 100\,\mathrm{\mu W}$ and $P_{\mathrm{photon}} = 1\,\mathrm{mW}$, respectively. Here $\widetilde{M}=0$ is assumed. (b) The intensity-normalized nonlinear radiation rate to the air of different OAM orders for microcavities of various radii. There is a trade-off between the maximal OAM order and radiation efficiency.}
\label{Fig3}
\end{figure}

Figure~\ref{Fig3}(a) shows the calculated radiation efficiency $\eta$ as a function of the photonic ($Q_1$) and phononic ($Q_2$) quality factors, assuming critical coupling ($\kappa_{\text{ex}}=\kappa_{\text{int}}$) and a fixed input microwave power $P_{\text{phonon}}=100\,\mu \text{W}$. In the regime where $Q_2$ is significantly lower than $Q_1$, the phonon-driven nonlinear radiative decay rate is negligible ($\kappa_\mathrm{nl}\ll\kappa_\mathrm{int}$), then $\eta\approx \kappa_\mathrm{nl}/\kappa_\mathrm{int}$ scales proportionally with the product $Q_{1}Q_{2}$ [Fig.~\ref{Fig3}(a)], a behavior that is consistent with the prediction from Eq.~\eqref{eq:radiationfield}. However, as the $Q_{1,2}$ increases further, the backaction from Brillouin NOR modifies the response of the optical mode, i.e., effectively shift the cavity-waveguide coupling from the critical coupling toward the under-coupled regime, leading to the saturation and reducing of the overall efficiency, as evident in the upper right corner of Fig.~\ref{Fig3}(a). The maximum $\eta=50\%$ is achieved at $\kappa_{\text{nl}} = 2\kappa_\text{int}$. Therefore, when the practical phonon drive amplitude is strong enough, we should optimize $\kappa_\mathrm{ex}$ in the over-coupling condition $\kappa_{\text{ex}}>\kappa_{\text{int}}$ for higher $\eta$. Remarkably, our model predicts efficiencies exceeding $25\%$ with experimentally feasible parameters, such as $Q_1\approx10^6$, $Q_2\approx5000$~\cite{Xu2025b,Yang2025a,Yang2025b} in lithium niobate on sapphire platform. We anticipate that similar performance can be achieved with other material platforms for integrated photonics~\cite{Liu2022}, including silicon~\cite{Chen2025silicon} and gallium nitride~\cite{Fu2019,Zhang2025,Xu2025a}. Figure~\ref{Fig3}(b) examines the dependence of $\kappa_2$ on OAM order-dependence, revealing a sharp cut-off for the nonlinear Brillouin NOR interactions that align with the phase-matching constraints predicted in Fig.~\ref{Fig2}(e). Comparisons across different ring radii $R$ demonstrate an inversely proportional relationship $\kappa_2\propto 1/R$. This highlights a key design trade-off in vortex beam generation: smaller rings yield higher $\eta$, whereas larger rings enable access to higher OAM orders, as shown in Fig.~\ref{Fig2}(e).

\smallskip{}
\noindent
\textit{Discussion.--} A key advantage of our vortex beam emitter lies in its capability of rapid reconfiguration. The OAM order of the emitted beam is determined by the OAM of the drive phononic mode, which is directly selected via the input microwave frequency. Given that the free spectral range ($\Delta$) of phononic WGMs in such cavities typically falls in the MHz range~\cite{Xu2022}, switching between adjacent OAM orders requires only a small MHz-level adjustment in the microwave drive. This allows for high-speed OAM reconfiguration solely through electrical control, compared with the optical wavelength switching with static nanostructures~\cite{caiscience} and the slow thermo-optic modulators in conventional devices~\cite{zhao2025all}.

\begin{figure}
\begin{centering}
\includegraphics[width=0.8\columnwidth]{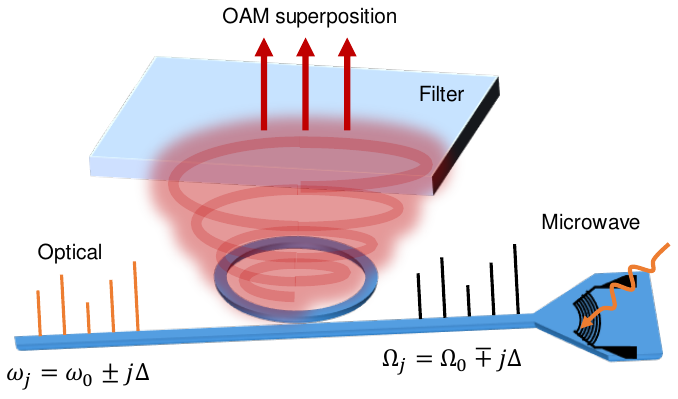}
\par\end{centering}
\caption{Universal synthesis of OAM superpositions. A single photonic WGM is excited with multi-tone optical field, while a group of phononic WGMs are excited via corresponding multi-tone microwave drive. A narrow-band spectral filter isolates the desired frequency-degenerate OAM superposition.}
\label{Fig4}
\end{figure}

Another crucial requirement for advanced coherent information processing is the generation of arbitrary OAM superpositions, $|\Psi\rangle=\sum c_j |\widetilde M_j\rangle$, at a fixed optical frequency. It is exceptionally challenging to create complex and spatially varying phase profiles required for such tasks by conventional on-chip OAM devices. Our Brillouin NOR scheme, in contrast, leverages the direct mapping between microwave frequency and OAM order through phononic dispersion relations, to provide an elegant and powerful solution. Figure~\ref{Fig4} illustrates the principle to generate a superposition of target OAM orders $\{\widetilde M_j\}$. We excite a target photonic WGM (order $m$) using multi-tone optical pump and simultaneously excite multiple phononic WGMs (one for each target OAM order, $\{-M_j\}$) by corresponding multi-tone microwave drive. Each microwave frequency tone at $\Omega_j=\Omega_0+j\Delta$ selectively excites a specific phononic WGM $M_j$, with the amplitude and phase of each OAM component in the final superposition are directly controlled by the amplitude $\varepsilon_j$ and phase $\theta_j$ of the corresponding microwave tone. It is worth noting that all optical tones can respond to all microwave drive tones, and the optical pump tones are tailored such that $\omega_j = \omega_0 - j \Delta$, ensuring SFR process yields a common output frequency $\omega_{\rm out} = \omega_j + \Omega_j = \omega_0 + \Omega_0$ for all $j$, and then an optical narrowband filter can filter out the target OAM superposition.

\smallskip{}
\noindent
\textit{Conclusion.--}
In conclusion, we have introduced and validated Brillouin NOR for generating structured light. This approach leverages microwave-driven acoustic waves as a reconfigurable, dynamic grating within a simple microring resonator. It eliminates the reliance on fixed, nanofabricated structures and unlocks a suite of powerful capabilities. We have shown that the OAM of the emitted beam can be rapidly tuned by the microwave frequency. More profoundly, we have demonstrated that arbitrary, high-dimensional OAM superpositions can be synthesized at a single wavelength by simply programming the spectrum of the microwave drive. With a predicted efficiency exceeding 25\% in standard photonic platforms, this method is not only versatile but also highly practical. Our work fundamentally transforms chip-to-free-space interfaces from static, lithographically-defined optics to dynamic, all-electronically controlled acousto-optic systems. This opens up exciting new avenues for advanced applications in quantum communication, optical trapping, and sensing that require programmable, on-demand structured light. Furthermore, the reversal process of Brillouin NOR suggests that our device could function not only as an emitter but also as a compact, mode-selective detector, enabling analysis of complex optical wavefronts.

\smallskip{}
\begin{acknowledgments}
We acknowledge Y.-L. Zhang for helpful discussion. This work was funded by the National Natural Science Foundation of China (Grant No.12374361, 12293053, 92265210, 123B2068, 12504454, 92165209, 92365301, and 92565301). This work is also supported by the Fundamental Research Funds for the Central Universities, the USTC Research Funds of the Double First-Class Initiative, the supercomputing system in the Supercomputing Center of USTC the USTC Center for Micro and Nanoscale Research and Fabrication. 
\end{acknowledgments}


\begin{thebibliography}{49}%
	\makeatletter
	\providecommand \@ifxundefined [1]{%
		\@ifx{#1\undefined}
	}%
	\providecommand \@ifnum [1]{%
		\ifnum #1\expandafter \@firstoftwo
		\else \expandafter \@secondoftwo
		\fi
	}%
	\providecommand \@ifx [1]{%
		\ifx #1\expandafter \@firstoftwo
		\else \expandafter \@secondoftwo
		\fi
	}%
	\providecommand \natexlab [1]{#1}%
	\providecommand \enquote  [1]{``#1''}%
	\providecommand \bibnamefont  [1]{#1}%
	\providecommand \bibfnamefont [1]{#1}%
	\providecommand \citenamefont [1]{#1}%
	\providecommand \href@noop [0]{\@secondoftwo}%
	\providecommand \href [0]{\begingroup \@sanitize@url \@href}%
	\providecommand \@href[1]{\@@startlink{#1}\@@href}%
	\providecommand \@@href[1]{\endgroup#1\@@endlink}%
	\providecommand \@sanitize@url [0]{\catcode `\\12\catcode `\$12\catcode
		`\&12\catcode `\#12\catcode `\^12\catcode `\_12\catcode `\%12\relax}%
	\providecommand \@@startlink[1]{}%
	\providecommand \@@endlink[0]{}%
	\providecommand \url  [0]{\begingroup\@sanitize@url \@url }%
	\providecommand \@url [1]{\endgroup\@href {#1}{\urlprefix }}%
	\providecommand \urlprefix  [0]{URL }%
	\providecommand \Eprint [0]{\href }%
	\providecommand \doibase [0]{http://dx.doi.org/}%
	\providecommand \selectlanguage [0]{\@gobble}%
	\providecommand \bibinfo  [0]{\@secondoftwo}%
	\providecommand \bibfield  [0]{\@secondoftwo}%
	\providecommand \translation [1]{[#1]}%
	\providecommand \BibitemOpen [0]{}%
	\providecommand \bibitemStop [0]{}%
	\providecommand \bibitemNoStop [0]{.\EOS\space}%
	\providecommand \EOS [0]{\spacefactor3000\relax}%
	\providecommand \BibitemShut  [1]{\csname bibitem#1\endcsname}%
	\let\auto@bib@innerbib\@empty
	\bibitem [{\citenamefont {Wang}\ \emph {et~al.}(2012)\citenamefont {Wang},
		\citenamefont {Yang}, \citenamefont {Fazal}, \citenamefont {Ahmed},
		\citenamefont {Yan}, \citenamefont {Huang}, \citenamefont {Ren},
		\citenamefont {Yue}, \citenamefont {Dolinar}, \citenamefont {Tur} \emph
		{et~al.}}]{wang2012terabit}%
	\BibitemOpen
	\bibfield  {author} {\bibinfo {author} {\bibfnamefont {J.}~\bibnamefont
			{Wang}}, \bibinfo {author} {\bibfnamefont {J.-Y.}\ \bibnamefont {Yang}},
		\bibinfo {author} {\bibfnamefont {I.~M.}\ \bibnamefont {Fazal}}, \bibinfo
		{author} {\bibfnamefont {N.}~\bibnamefont {Ahmed}}, \bibinfo {author}
		{\bibfnamefont {Y.}~\bibnamefont {Yan}}, \bibinfo {author} {\bibfnamefont
			{H.}~\bibnamefont {Huang}}, \bibinfo {author} {\bibfnamefont
			{Y.}~\bibnamefont {Ren}}, \bibinfo {author} {\bibfnamefont {Y.}~\bibnamefont
			{Yue}}, \bibinfo {author} {\bibfnamefont {S.}~\bibnamefont {Dolinar}},
		\bibinfo {author} {\bibfnamefont {M.}~\bibnamefont {Tur}},  \emph {et~al.},\
	}\bibfield  {title} {\enquote {\bibinfo {title} {Terabit free-space data
				transmission employing orbital angular momentum multiplexing},}\ }\href
	{https://doi.org/10.1038/nphoton.2012.138} {\bibfield  {journal} {\bibinfo
			{journal} {Nature Photonics}\ }\textbf {\bibinfo {volume} {6}},\ \bibinfo
		{pages} {488} (\bibinfo {year} {2012})}\BibitemShut {NoStop}%
	\bibitem [{\citenamefont {Molina-Terriza}\ \emph {et~al.}(2007)\citenamefont
		{Molina-Terriza}, \citenamefont {Torres},\ and\ \citenamefont
		{Torner}}]{molina2007twisted}%
	\BibitemOpen
	\bibfield  {author} {\bibinfo {author} {\bibfnamefont {G.}~\bibnamefont
			{Molina-Terriza}}, \bibinfo {author} {\bibfnamefont {J.~P.}\ \bibnamefont
			{Torres}}, \ and\ \bibinfo {author} {\bibfnamefont {L.}~\bibnamefont
			{Torner}},\ }\bibfield  {title} {\enquote {\bibinfo {title} {Twisted
				photons},}\ }\href {https://doi.org/10.1038/nphys607} {\bibfield  {journal}
		{\bibinfo  {journal} {Nature Physics}\ }\textbf {\bibinfo {volume} {3}},\
		\bibinfo {pages} {305} (\bibinfo {year} {2007})}\BibitemShut {NoStop}%
	\bibitem [{\citenamefont {Fickler}\ \emph {et~al.}(2012)\citenamefont
		{Fickler}, \citenamefont {Lapkiewicz}, \citenamefont {Plick}, \citenamefont
		{Krenn}, \citenamefont {Schaeff}, \citenamefont {Ramelow},\ and\
		\citenamefont {Zeilinger}}]{fickler2012quantum}%
	\BibitemOpen
	\bibfield  {author} {\bibinfo {author} {\bibfnamefont {R.}~\bibnamefont
			{Fickler}}, \bibinfo {author} {\bibfnamefont {R.}~\bibnamefont {Lapkiewicz}},
		\bibinfo {author} {\bibfnamefont {W.~N.}\ \bibnamefont {Plick}}, \bibinfo
		{author} {\bibfnamefont {M.}~\bibnamefont {Krenn}}, \bibinfo {author}
		{\bibfnamefont {C.}~\bibnamefont {Schaeff}}, \bibinfo {author} {\bibfnamefont
			{S.}~\bibnamefont {Ramelow}}, \ and\ \bibinfo {author} {\bibfnamefont
			{A.}~\bibnamefont {Zeilinger}},\ }\bibfield  {title} {\enquote {\bibinfo
			{title} {Quantum entanglement of high angular momenta},}\ }\href
	{https://doi.org/10.1126/science.1227193} {\bibfield  {journal} {\bibinfo
			{journal} {Science}\ }\textbf {\bibinfo {volume} {338}},\ \bibinfo {pages}
		{640} (\bibinfo {year} {2012})}\BibitemShut {NoStop}%
	\bibitem [{\citenamefont {Fickler}\ \emph {et~al.}(2016)\citenamefont
		{Fickler}, \citenamefont {Campbell}, \citenamefont {Buchler}, \citenamefont
		{Lam},\ and\ \citenamefont {Zeilinger}}]{fickler2016quantum}%
	\BibitemOpen
	\bibfield  {author} {\bibinfo {author} {\bibfnamefont {R.}~\bibnamefont
			{Fickler}}, \bibinfo {author} {\bibfnamefont {G.}~\bibnamefont {Campbell}},
		\bibinfo {author} {\bibfnamefont {B.}~\bibnamefont {Buchler}}, \bibinfo
		{author} {\bibfnamefont {P.~K.}\ \bibnamefont {Lam}}, \ and\ \bibinfo
		{author} {\bibfnamefont {A.}~\bibnamefont {Zeilinger}},\ }\bibfield  {title}
	{\enquote {\bibinfo {title} {Quantum entanglement of angular momentum states
				with quantum numbers up to 10,010},}\ }\href
	{https://doi.org/10.1073/pnas.1616889113} {\bibfield  {journal} {\bibinfo
			{journal} {Proceedings of the National Academy of Sciences}\ }\textbf
		{\bibinfo {volume} {113}},\ \bibinfo {pages} {13642} (\bibinfo {year}
		{2016})}\BibitemShut {NoStop}%
	\bibitem [{\citenamefont {Stav}\ \emph {et~al.}(2018)\citenamefont {Stav},
		\citenamefont {Faerman}, \citenamefont {Maguid}, \citenamefont {Oren},
		\citenamefont {Kleiner}, \citenamefont {Hasman},\ and\ \citenamefont
		{Segev}}]{stav2018quantum}%
	\BibitemOpen
	\bibfield  {author} {\bibinfo {author} {\bibfnamefont {T.}~\bibnamefont
			{Stav}}, \bibinfo {author} {\bibfnamefont {A.}~\bibnamefont {Faerman}},
		\bibinfo {author} {\bibfnamefont {E.}~\bibnamefont {Maguid}}, \bibinfo
		{author} {\bibfnamefont {D.}~\bibnamefont {Oren}}, \bibinfo {author}
		{\bibfnamefont {V.}~\bibnamefont {Kleiner}}, \bibinfo {author} {\bibfnamefont
			{E.}~\bibnamefont {Hasman}}, \ and\ \bibinfo {author} {\bibfnamefont
			{M.}~\bibnamefont {Segev}},\ }\bibfield  {title} {\enquote {\bibinfo {title}
			{Quantum entanglement of the spin and orbital angular momentum of photons
				using metamaterials},}\ }\href {https://doi.org/10.1126/science.aat9042}
	{\bibfield  {journal} {\bibinfo  {journal} {Science}\ }\textbf {\bibinfo
			{volume} {361}},\ \bibinfo {pages} {1101} (\bibinfo {year}
		{2018})}\BibitemShut {NoStop}%
	\bibitem [{\citenamefont {Erhard}\ \emph {et~al.}(2020)\citenamefont {Erhard},
		\citenamefont {Krenn},\ and\ \citenamefont {Zeilinger}}]{erhard2020advances}%
	\BibitemOpen
	\bibfield  {author} {\bibinfo {author} {\bibfnamefont {M.}~\bibnamefont
			{Erhard}}, \bibinfo {author} {\bibfnamefont {M.}~\bibnamefont {Krenn}}, \
		and\ \bibinfo {author} {\bibfnamefont {A.}~\bibnamefont {Zeilinger}},\
	}\bibfield  {title} {\enquote {\bibinfo {title} {Advances in high-dimensional
				quantum entanglement},}\ }\href {https://doi.org/10.1038/s42254-020-0193-5}
	{\bibfield  {journal} {\bibinfo  {journal} {Nature Reviews Physics}\ }\textbf
		{\bibinfo {volume} {2}},\ \bibinfo {pages} {365} (\bibinfo {year}
		{2020})}\BibitemShut {NoStop}%
	\bibitem [{\citenamefont {Ding}\ \emph {et~al.}(2015)\citenamefont {Ding},
		\citenamefont {Zhang}, \citenamefont {Zhou}, \citenamefont {Shi},
		\citenamefont {Xiang}, \citenamefont {Wang}, \citenamefont {Jiang},
		\citenamefont {Shi},\ and\ \citenamefont {Guo}}]{ding2015quantum}%
	\BibitemOpen
	\bibfield  {author} {\bibinfo {author} {\bibfnamefont {D.-S.}\ \bibnamefont
			{Ding}}, \bibinfo {author} {\bibfnamefont {W.}~\bibnamefont {Zhang}},
		\bibinfo {author} {\bibfnamefont {Z.-Y.}\ \bibnamefont {Zhou}}, \bibinfo
		{author} {\bibfnamefont {S.}~\bibnamefont {Shi}}, \bibinfo {author}
		{\bibfnamefont {G.-Y.}\ \bibnamefont {Xiang}}, \bibinfo {author}
		{\bibfnamefont {X.-S.}\ \bibnamefont {Wang}}, \bibinfo {author}
		{\bibfnamefont {Y.-K.}\ \bibnamefont {Jiang}}, \bibinfo {author}
		{\bibfnamefont {B.-S.}\ \bibnamefont {Shi}}, \ and\ \bibinfo {author}
		{\bibfnamefont {G.-C.}\ \bibnamefont {Guo}},\ }\bibfield  {title} {\enquote
		{\bibinfo {title} {Quantum storage of orbital angular momentum entanglement
				in an atomic ensemble},}\ }\href
	{https://doi.org/10.1103/PhysRevLett.114.050502} {\bibfield  {journal}
		{\bibinfo  {journal} {Physical Review Letters}\ }\textbf {\bibinfo {volume}
			{114}},\ \bibinfo {pages} {050502} (\bibinfo {year} {2015})}\BibitemShut
	{NoStop}%
	\bibitem [{\citenamefont {Mirhosseini}\ \emph {et~al.}(2015)\citenamefont
		{Mirhosseini}, \citenamefont {Maga{\~n}a-Loaiza}, \citenamefont
		{O’Sullivan}, \citenamefont {Rodenburg}, \citenamefont {Malik},
		\citenamefont {Lavery}, \citenamefont {Padgett}, \citenamefont {Gauthier},\
		and\ \citenamefont {Boyd}}]{mirhosseini2015high}%
	\BibitemOpen
	\bibfield  {author} {\bibinfo {author} {\bibfnamefont {M.}~\bibnamefont
			{Mirhosseini}}, \bibinfo {author} {\bibfnamefont {O.~S.}\ \bibnamefont
			{Maga{\~n}a-Loaiza}}, \bibinfo {author} {\bibfnamefont {M.~N.}\ \bibnamefont
			{O’Sullivan}}, \bibinfo {author} {\bibfnamefont {B.}~\bibnamefont
			{Rodenburg}}, \bibinfo {author} {\bibfnamefont {M.}~\bibnamefont {Malik}},
		\bibinfo {author} {\bibfnamefont {M.~P.}\ \bibnamefont {Lavery}}, \bibinfo
		{author} {\bibfnamefont {M.~J.}\ \bibnamefont {Padgett}}, \bibinfo {author}
		{\bibfnamefont {D.~J.}\ \bibnamefont {Gauthier}}, \ and\ \bibinfo {author}
		{\bibfnamefont {R.~W.}\ \bibnamefont {Boyd}},\ }\bibfield  {title} {\enquote
		{\bibinfo {title} {High-dimensional quantum cryptography with twisted
				light},}\ }\href {https://doi.org/10.1088/1367-2630/17/3/033033} {\bibfield
		{journal} {\bibinfo  {journal} {New Journal of Physics}\ }\textbf {\bibinfo
			{volume} {17}},\ \bibinfo {pages} {033033} (\bibinfo {year}
		{2015})}\BibitemShut {NoStop}%
	\bibitem [{\citenamefont {Grier}(2003)}]{grier2003revolution}%
	\BibitemOpen
	\bibfield  {author} {\bibinfo {author} {\bibfnamefont {D.~G.}\ \bibnamefont
			{Grier}},\ }\bibfield  {title} {\enquote {\bibinfo {title} {A revolution in
				optical manipulation},}\ }\href {https://doi.org/10.1038/nature01935}
	{\bibfield  {journal} {\bibinfo  {journal} {Nature}\ }\textbf {\bibinfo
			{volume} {424}},\ \bibinfo {pages} {810} (\bibinfo {year}
		{2003})}\BibitemShut {NoStop}%
	\bibitem [{\citenamefont {Padgett}\ and\ \citenamefont
		{Bowman}(2011)}]{padgett2011tweezers}%
	\BibitemOpen
	\bibfield  {author} {\bibinfo {author} {\bibfnamefont {M.}~\bibnamefont
			{Padgett}}\ and\ \bibinfo {author} {\bibfnamefont {R.}~\bibnamefont
			{Bowman}},\ }\bibfield  {title} {\enquote {\bibinfo {title} {Tweezers with a
				twist},}\ }\href {https://doi.org/10.1038/nphoton.2011.81} {\bibfield
		{journal} {\bibinfo  {journal} {Nature Photonics}\ }\textbf {\bibinfo
			{volume} {5}},\ \bibinfo {pages} {343} (\bibinfo {year} {2011})}\BibitemShut
	{NoStop}%
	\bibitem [{\citenamefont {Paterson}\ \emph {et~al.}(2001)\citenamefont
		{Paterson}, \citenamefont {MacDonald}, \citenamefont {Arlt}, \citenamefont
		{Sibbett}, \citenamefont {Bryant},\ and\ \citenamefont
		{Dholakia}}]{paterson2001controlled}%
	\BibitemOpen
	\bibfield  {author} {\bibinfo {author} {\bibfnamefont {L.}~\bibnamefont
			{Paterson}}, \bibinfo {author} {\bibfnamefont {M.~P.}\ \bibnamefont
			{MacDonald}}, \bibinfo {author} {\bibfnamefont {J.}~\bibnamefont {Arlt}},
		\bibinfo {author} {\bibfnamefont {W.}~\bibnamefont {Sibbett}}, \bibinfo
		{author} {\bibfnamefont {P.~E.}\ \bibnamefont {Bryant}}, \ and\ \bibinfo
		{author} {\bibfnamefont {K.}~\bibnamefont {Dholakia}},\ }\bibfield  {title}
	{\enquote {\bibinfo {title} {Controlled rotation of optically trapped
				microscopic particles},}\ }\href {https://doi.org/10.1126/science.1058591}
	{\bibfield  {journal} {\bibinfo  {journal} {Science}\ }\textbf {\bibinfo
			{volume} {292}},\ \bibinfo {pages} {912} (\bibinfo {year}
		{2001})}\BibitemShut {NoStop}%
	\bibitem [{\citenamefont {Shen}\ \emph {et~al.}(2019)\citenamefont {Shen},
		\citenamefont {Wang}, \citenamefont {Xie}, \citenamefont {Min}, \citenamefont
		{Fu}, \citenamefont {Liu}, \citenamefont {Gong},\ and\ \citenamefont
		{Yuan}}]{shen2019optical}%
	\BibitemOpen
	\bibfield  {author} {\bibinfo {author} {\bibfnamefont {Y.}~\bibnamefont
			{Shen}}, \bibinfo {author} {\bibfnamefont {X.}~\bibnamefont {Wang}}, \bibinfo
		{author} {\bibfnamefont {Z.}~\bibnamefont {Xie}}, \bibinfo {author}
		{\bibfnamefont {C.}~\bibnamefont {Min}}, \bibinfo {author} {\bibfnamefont
			{X.}~\bibnamefont {Fu}}, \bibinfo {author} {\bibfnamefont {Q.}~\bibnamefont
			{Liu}}, \bibinfo {author} {\bibfnamefont {M.}~\bibnamefont {Gong}}, \ and\
		\bibinfo {author} {\bibfnamefont {X.}~\bibnamefont {Yuan}},\ }\bibfield
	{title} {\enquote {\bibinfo {title} {Optical vortices 30 years on: {OAM}
				manipulation from topological charge to multiple singularities},}\ }\href
	{https://doi.org/10.1038/s41377-019-0194-2} {\bibfield  {journal} {\bibinfo
			{journal} {Light: Science \& Applications}\ }\textbf {\bibinfo {volume}
			{8}},\ \bibinfo {pages} {90} (\bibinfo {year} {2019})}\BibitemShut {NoStop}%
	\bibitem [{\citenamefont {F{\"u}rhapter}\ \emph {et~al.}(2005)\citenamefont
		{F{\"u}rhapter}, \citenamefont {Jesacher}, \citenamefont {Bernet},\ and\
		\citenamefont {Ritsch-Marte}}]{furhapter2005spiral}%
	\BibitemOpen
	\bibfield  {author} {\bibinfo {author} {\bibfnamefont {S.}~\bibnamefont
			{F{\"u}rhapter}}, \bibinfo {author} {\bibfnamefont {A.}~\bibnamefont
			{Jesacher}}, \bibinfo {author} {\bibfnamefont {S.}~\bibnamefont {Bernet}}, \
		and\ \bibinfo {author} {\bibfnamefont {M.}~\bibnamefont {Ritsch-Marte}},\
	}\bibfield  {title} {\enquote {\bibinfo {title} {Spiral phase contrast
				imaging in microscopy},}\ }\href {https://doi.org/10.1364/OPEX.13.000689}
	{\bibfield  {journal} {\bibinfo  {journal} {Optics Express}\ }\textbf
		{\bibinfo {volume} {13}},\ \bibinfo {pages} {689} (\bibinfo {year}
		{2005})}\BibitemShut {NoStop}%
	\bibitem [{\citenamefont {Tamburini}\ \emph {et~al.}(2006)\citenamefont
		{Tamburini}, \citenamefont {Anzolin}, \citenamefont {Umbriaco}, \citenamefont
		{Bianchini},\ and\ \citenamefont {Barbieri}}]{PhysRevLett.97.163903}%
	\BibitemOpen
	\bibfield  {author} {\bibinfo {author} {\bibfnamefont {F.}~\bibnamefont
			{Tamburini}}, \bibinfo {author} {\bibfnamefont {G.}~\bibnamefont {Anzolin}},
		\bibinfo {author} {\bibfnamefont {G.}~\bibnamefont {Umbriaco}}, \bibinfo
		{author} {\bibfnamefont {A.}~\bibnamefont {Bianchini}}, \ and\ \bibinfo
		{author} {\bibfnamefont {C.}~\bibnamefont {Barbieri}},\ }\bibfield  {title}
	{\enquote {\bibinfo {title} {Overcoming the rayleigh criterion limit with
				optical vortices},}\ }\href {\doibase 10.1103/PhysRevLett.97.163903}
	{\bibfield  {journal} {\bibinfo  {journal} {Physical Review Letters}\
		}\textbf {\bibinfo {volume} {97}},\ \bibinfo {pages} {163903} (\bibinfo
		{year} {2006})}\BibitemShut {NoStop}%
	\bibitem [{\citenamefont {Yao}\ and\ \citenamefont
		{Padgett}(2011)}]{yao2011orbital}%
	\BibitemOpen
	\bibfield  {author} {\bibinfo {author} {\bibfnamefont {A.~M.}\ \bibnamefont
			{Yao}}\ and\ \bibinfo {author} {\bibfnamefont {M.~J.}\ \bibnamefont
			{Padgett}},\ }\bibfield  {title} {\enquote {\bibinfo {title} {Orbital angular
				momentum: origins, behavior and applications},}\ }\href
	{https://doi.org/10.1364/AOP.3.000161} {\bibfield  {journal} {\bibinfo
			{journal} {Advances in optics and photonics}\ }\textbf {\bibinfo {volume}
			{3}},\ \bibinfo {pages} {161} (\bibinfo {year} {2011})}\BibitemShut {NoStop}%
	\bibitem [{\citenamefont {Forbes}\ \emph {et~al.}(2021)\citenamefont {Forbes},
		\citenamefont {De~Oliveira},\ and\ \citenamefont
		{Dennis}}]{forbes2021structured}%
	\BibitemOpen
	\bibfield  {author} {\bibinfo {author} {\bibfnamefont {A.}~\bibnamefont
			{Forbes}}, \bibinfo {author} {\bibfnamefont {M.}~\bibnamefont {De~Oliveira}},
		\ and\ \bibinfo {author} {\bibfnamefont {M.~R.}\ \bibnamefont {Dennis}},\
	}\bibfield  {title} {\enquote {\bibinfo {title} {Structured light},}\ }\href
	{https://doi.org/10.1038/s41566-021-00780-4} {\bibfield  {journal} {\bibinfo
			{journal} {Nature Photonics}\ }\textbf {\bibinfo {volume} {15}},\ \bibinfo
		{pages} {253} (\bibinfo {year} {2021})}\BibitemShut {NoStop}%
	\bibitem [{\citenamefont {Chen}\ \emph
		{et~al.}(2021{\natexlab{a}})\citenamefont {Chen}, \citenamefont {Wan},\ and\
		\citenamefont {Zhan}}]{chen2021engineering}%
	\BibitemOpen
	\bibfield  {author} {\bibinfo {author} {\bibfnamefont {J.}~\bibnamefont
			{Chen}}, \bibinfo {author} {\bibfnamefont {C.}~\bibnamefont {Wan}}, \ and\
		\bibinfo {author} {\bibfnamefont {Q.}~\bibnamefont {Zhan}},\ }\bibfield
	{title} {\enquote {\bibinfo {title} {Engineering photonic angular momentum
				with structured light: a review},}\ }\href
	{https://doi.org/10.1117/1.AP.3.6.064001} {\bibfield  {journal} {\bibinfo
			{journal} {Advanced Photonics}\ }\textbf {\bibinfo {volume} {3}},\ \bibinfo
		{pages} {064001} (\bibinfo {year} {2021}{\natexlab{a}})}\BibitemShut
	{NoStop}%
	\bibitem [{\citenamefont {Yang}\ \emph
		{et~al.}(2021{\natexlab{a}})\citenamefont {Yang}, \citenamefont {Xie},
		\citenamefont {He}, \citenamefont {Zhang},\ and\ \citenamefont
		{Yuan}}]{yang2021perspective}%
	\BibitemOpen
	\bibfield  {author} {\bibinfo {author} {\bibfnamefont {H.}~\bibnamefont
			{Yang}}, \bibinfo {author} {\bibfnamefont {Z.}~\bibnamefont {Xie}}, \bibinfo
		{author} {\bibfnamefont {H.}~\bibnamefont {He}}, \bibinfo {author}
		{\bibfnamefont {Q.}~\bibnamefont {Zhang}}, \ and\ \bibinfo {author}
		{\bibfnamefont {X.}~\bibnamefont {Yuan}},\ }\bibfield  {title} {\enquote
		{\bibinfo {title} {A perspective on twisted light from on-chip devices},}\
	}\href {https://doi.org/10.1063/5.0060736} {\bibfield  {journal} {\bibinfo
			{journal} {APL Photonics}\ }\textbf {\bibinfo {volume} {6}},\ \bibinfo
		{pages} {110901} (\bibinfo {year} {2021}{\natexlab{a}})}\BibitemShut
	{NoStop}%
	\bibitem [{\citenamefont {Devlin}\ \emph {et~al.}(2017)\citenamefont {Devlin},
		\citenamefont {Ambrosio}, \citenamefont {Rubin}, \citenamefont {Mueller},\
		and\ \citenamefont {Capasso}}]{scienceOAMspin}%
	\BibitemOpen
	\bibfield  {author} {\bibinfo {author} {\bibfnamefont {R.~C.}\ \bibnamefont
			{Devlin}}, \bibinfo {author} {\bibfnamefont {A.}~\bibnamefont {Ambrosio}},
		\bibinfo {author} {\bibfnamefont {N.~A.}\ \bibnamefont {Rubin}}, \bibinfo
		{author} {\bibfnamefont {J.~P.~B.}\ \bibnamefont {Mueller}}, \ and\ \bibinfo
		{author} {\bibfnamefont {F.}~\bibnamefont {Capasso}},\ }\bibfield  {title}
	{\enquote {\bibinfo {title} {Arbitrary spin-to–orbital angular momentum
				conversion of light},}\ }\href {\doibase 10.1126/science.aao5392} {\bibfield
		{journal} {\bibinfo  {journal} {Science}\ }\textbf {\bibinfo {volume}
			{358}},\ \bibinfo {pages} {896} (\bibinfo {year} {2017})}\BibitemShut
	{NoStop}%
	\bibitem [{\citenamefont {Sroor}\ \emph {et~al.}(2020)\citenamefont {Sroor},
		\citenamefont {Huang}, \citenamefont {Sephton}, \citenamefont {Naidoo},
		\citenamefont {Vall{\'e}s}, \citenamefont {Ginis}, \citenamefont {Qiu},
		\citenamefont {Ambrosio}, \citenamefont {Capasso},\ and\ \citenamefont
		{Forbes}}]{sroor2020high}%
	\BibitemOpen
	\bibfield  {author} {\bibinfo {author} {\bibfnamefont {H.}~\bibnamefont
			{Sroor}}, \bibinfo {author} {\bibfnamefont {Y.-W.}\ \bibnamefont {Huang}},
		\bibinfo {author} {\bibfnamefont {B.}~\bibnamefont {Sephton}}, \bibinfo
		{author} {\bibfnamefont {D.}~\bibnamefont {Naidoo}}, \bibinfo {author}
		{\bibfnamefont {A.}~\bibnamefont {Vall{\'e}s}}, \bibinfo {author}
		{\bibfnamefont {V.}~\bibnamefont {Ginis}}, \bibinfo {author} {\bibfnamefont
			{C.-W.}\ \bibnamefont {Qiu}}, \bibinfo {author} {\bibfnamefont
			{A.}~\bibnamefont {Ambrosio}}, \bibinfo {author} {\bibfnamefont
			{F.}~\bibnamefont {Capasso}}, \ and\ \bibinfo {author} {\bibfnamefont
			{A.}~\bibnamefont {Forbes}},\ }\bibfield  {title} {\enquote {\bibinfo {title}
			{High-purity orbital angular momentum states from a visible metasurface
				laser},}\ }\href {https://doi.org/10.1038/s41566-020-0623-z} {\bibfield
		{journal} {\bibinfo  {journal} {Nature Photonics}\ }\textbf {\bibinfo
			{volume} {14}},\ \bibinfo {pages} {498} (\bibinfo {year} {2020})}\BibitemShut
	{NoStop}%
	\bibitem [{\citenamefont {Yang}\ \emph
		{et~al.}(2021{\natexlab{b}})\citenamefont {Yang}, \citenamefont {Xie},
		\citenamefont {Li}, \citenamefont {Ou}, \citenamefont {Yu}, \citenamefont
		{He}, \citenamefont {Wang},\ and\ \citenamefont {Yuan}}]{Yang:21}%
	\BibitemOpen
	\bibfield  {author} {\bibinfo {author} {\bibfnamefont {H.}~\bibnamefont
			{Yang}}, \bibinfo {author} {\bibfnamefont {Z.}~\bibnamefont {Xie}}, \bibinfo
		{author} {\bibfnamefont {G.}~\bibnamefont {Li}}, \bibinfo {author}
		{\bibfnamefont {K.}~\bibnamefont {Ou}}, \bibinfo {author} {\bibfnamefont
			{F.}~\bibnamefont {Yu}}, \bibinfo {author} {\bibfnamefont {H.}~\bibnamefont
			{He}}, \bibinfo {author} {\bibfnamefont {H.}~\bibnamefont {Wang}}, \ and\
		\bibinfo {author} {\bibfnamefont {X.}~\bibnamefont {Yuan}},\ }\bibfield
	{title} {\enquote {\bibinfo {title} {All-dielectric metasurface for fully
				resolving arbitrary beams on a higher-order {P}oincar\'{e} sphere},}\ }\href
	{\doibase 10.1364/PRJ.411503} {\bibfield  {journal} {\bibinfo  {journal}
			{Photonics Research}\ }\textbf {\bibinfo {volume} {9}},\ \bibinfo {pages}
		{331} (\bibinfo {year} {2021}{\natexlab{b}})}\BibitemShut {NoStop}%
	\bibitem [{\citenamefont {Liang}\ \emph {et~al.}(2014)\citenamefont {Liang},
		\citenamefont {Wu}, \citenamefont {Huang},\ and\ \citenamefont
		{Huang}}]{C4NR03606A}%
	\BibitemOpen
	\bibfield  {author} {\bibinfo {author} {\bibfnamefont {Y.}~\bibnamefont
			{Liang}}, \bibinfo {author} {\bibfnamefont {H.~W.}\ \bibnamefont {Wu}},
		\bibinfo {author} {\bibfnamefont {B.~J.}\ \bibnamefont {Huang}}, \ and\
		\bibinfo {author} {\bibfnamefont {X.~G.}\ \bibnamefont {Huang}},\ }\bibfield
	{title} {\enquote {\bibinfo {title} {Light beams with selective angular
				momentum generated by hybrid plasmonic waveguides},}\ }\href {\doibase
		10.1039/C4NR03606A} {\bibfield  {journal} {\bibinfo  {journal} {Nanoscale}\
		}\textbf {\bibinfo {volume} {6}},\ \bibinfo {pages} {12360} (\bibinfo {year}
		{2014})}\BibitemShut {NoStop}%
	\bibitem [{\citenamefont {Liu}\ \emph {et~al.}(2016)\citenamefont {Liu},
		\citenamefont {Zou}, \citenamefont {Ren}, \citenamefont {Wang},\ and\
		\citenamefont {Guo}}]{liu2016chip}%
	\BibitemOpen
	\bibfield  {author} {\bibinfo {author} {\bibfnamefont {A.}~\bibnamefont
			{Liu}}, \bibinfo {author} {\bibfnamefont {C.-L.}\ \bibnamefont {Zou}},
		\bibinfo {author} {\bibfnamefont {X.}~\bibnamefont {Ren}}, \bibinfo {author}
		{\bibfnamefont {Q.}~\bibnamefont {Wang}}, \ and\ \bibinfo {author}
		{\bibfnamefont {G.-C.}\ \bibnamefont {Guo}},\ }\bibfield  {title} {\enquote
		{\bibinfo {title} {On-chip generation and control of the vortex beam},}\
	}\href {https://doi.org/10.1063/1.4948519} {\bibfield  {journal} {\bibinfo
			{journal} {Applied Physics Letters}\ }\textbf {\bibinfo {volume} {108}},\
		\bibinfo {pages} {181103} (\bibinfo {year} {2016})}\BibitemShut {NoStop}%
	\bibitem [{\citenamefont {Cai}\ \emph {et~al.}(2012)\citenamefont {Cai},
		\citenamefont {Wang}, \citenamefont {Strain}, \citenamefont {Johnson-Morris},
		\citenamefont {Zhu}, \citenamefont {Sorel}, \citenamefont {O'Brien},
		\citenamefont {Thompson},\ and\ \citenamefont {Yu}}]{caiscience}%
	\BibitemOpen
	\bibfield  {author} {\bibinfo {author} {\bibfnamefont {X.}~\bibnamefont
			{Cai}}, \bibinfo {author} {\bibfnamefont {J.}~\bibnamefont {Wang}}, \bibinfo
		{author} {\bibfnamefont {M.~J.}\ \bibnamefont {Strain}}, \bibinfo {author}
		{\bibfnamefont {B.}~\bibnamefont {Johnson-Morris}}, \bibinfo {author}
		{\bibfnamefont {J.}~\bibnamefont {Zhu}}, \bibinfo {author} {\bibfnamefont
			{M.}~\bibnamefont {Sorel}}, \bibinfo {author} {\bibfnamefont {J.~L.}\
			\bibnamefont {O'Brien}}, \bibinfo {author} {\bibfnamefont {M.~G.}\
			\bibnamefont {Thompson}}, \ and\ \bibinfo {author} {\bibfnamefont
			{S.}~\bibnamefont {Yu}},\ }\bibfield  {title} {\enquote {\bibinfo {title}
			{Integrated compact optical vortex beam emitters},}\ }\href {\doibase
		10.1126/science.1226528} {\bibfield  {journal} {\bibinfo  {journal}
			{Science}\ }\textbf {\bibinfo {volume} {338}},\ \bibinfo {pages} {363}
		(\bibinfo {year} {2012})}\BibitemShut {NoStop}%
	\bibitem [{\citenamefont {Zhao}\ \emph {et~al.}(2025)\citenamefont {Zhao},
		\citenamefont {Yi}, \citenamefont {Huang}, \citenamefont {Liu}, \citenamefont
		{Wang}, \citenamefont {Shi}, \citenamefont {Ma}, \citenamefont {Forbes},\
		and\ \citenamefont {Dai}}]{zhao2025all}%
	\BibitemOpen
	\bibfield  {author} {\bibinfo {author} {\bibfnamefont {W.}~\bibnamefont
			{Zhao}}, \bibinfo {author} {\bibfnamefont {X.}~\bibnamefont {Yi}}, \bibinfo
		{author} {\bibfnamefont {J.}~\bibnamefont {Huang}}, \bibinfo {author}
		{\bibfnamefont {R.}~\bibnamefont {Liu}}, \bibinfo {author} {\bibfnamefont
			{J.}~\bibnamefont {Wang}}, \bibinfo {author} {\bibfnamefont {Y.}~\bibnamefont
			{Shi}}, \bibinfo {author} {\bibfnamefont {Y.}~\bibnamefont {Ma}}, \bibinfo
		{author} {\bibfnamefont {A.}~\bibnamefont {Forbes}}, \ and\ \bibinfo {author}
		{\bibfnamefont {D.}~\bibnamefont {Dai}},\ }\bibfield  {title} {\enquote
		{\bibinfo {title} {All-on-chip reconfigurable generation of scalar and
				vectorial orbital angular momentum beams},}\ }\href
	{https://doi.org/10.1038/s41377-025-01899-7} {\bibfield  {journal} {\bibinfo
			{journal} {Light: Science \& Applications}\ }\textbf {\bibinfo {volume}
			{14}},\ \bibinfo {pages} {227} (\bibinfo {year} {2025})}\BibitemShut
	{NoStop}%
	\bibitem [{\citenamefont {Strekalov}\ \emph {et~al.}(2016)\citenamefont
		{Strekalov}, \citenamefont {Marquardt}, \citenamefont {Matsko}, \citenamefont
		{Schwefel},\ and\ \citenamefont {Leuchs}}]{strekalov2016nonlinear}%
	\BibitemOpen
	\bibfield  {author} {\bibinfo {author} {\bibfnamefont {D.~V.}\ \bibnamefont
			{Strekalov}}, \bibinfo {author} {\bibfnamefont {C.}~\bibnamefont
			{Marquardt}}, \bibinfo {author} {\bibfnamefont {A.~B.}\ \bibnamefont
			{Matsko}}, \bibinfo {author} {\bibfnamefont {H.~G.}\ \bibnamefont
			{Schwefel}}, \ and\ \bibinfo {author} {\bibfnamefont {G.}~\bibnamefont
			{Leuchs}},\ }\bibfield  {title} {\enquote {\bibinfo {title} {Nonlinear and
				quantum optics with whispering gallery resonators},}\ }\href
	{https://doi.org/10.1088/2040-8978/18/12/123002} {\bibfield  {journal}
		{\bibinfo  {journal} {Journal of Optics}\ }\textbf {\bibinfo {volume} {18}},\
		\bibinfo {pages} {123002} (\bibinfo {year} {2016})}\BibitemShut {NoStop}%
	\bibitem [{\citenamefont {Miao}\ \emph {et~al.}(2016)\citenamefont {Miao},
		\citenamefont {Zhang}, \citenamefont {Sun}, \citenamefont {Walasik},
		\citenamefont {Longhi}, \citenamefont {Litchinitser},\ and\ \citenamefont
		{Feng}}]{miao2016orbital}%
	\BibitemOpen
	\bibfield  {author} {\bibinfo {author} {\bibfnamefont {P.}~\bibnamefont
			{Miao}}, \bibinfo {author} {\bibfnamefont {Z.}~\bibnamefont {Zhang}},
		\bibinfo {author} {\bibfnamefont {J.}~\bibnamefont {Sun}}, \bibinfo {author}
		{\bibfnamefont {W.}~\bibnamefont {Walasik}}, \bibinfo {author} {\bibfnamefont
			{S.}~\bibnamefont {Longhi}}, \bibinfo {author} {\bibfnamefont {N.~M.}\
			\bibnamefont {Litchinitser}}, \ and\ \bibinfo {author} {\bibfnamefont
			{L.}~\bibnamefont {Feng}},\ }\bibfield  {title} {\enquote {\bibinfo {title}
			{Orbital angular momentum microlaser},}\ }\href
	{https://doi.org/10.1126/science.aaf8533} {\bibfield  {journal} {\bibinfo
			{journal} {Science}\ }\textbf {\bibinfo {volume} {353}},\ \bibinfo {pages}
		{464} (\bibinfo {year} {2016})}\BibitemShut {NoStop}%
	\bibitem [{\citenamefont {Chen}\ \emph
		{et~al.}(2021{\natexlab{b}})\citenamefont {Chen}, \citenamefont {Wei},
		\citenamefont {Zhao}, \citenamefont {Liu}, \citenamefont {Su}, \citenamefont
		{Yao}, \citenamefont {Yu}, \citenamefont {Liu},\ and\ \citenamefont
		{Wang}}]{chen2021bright}%
	\BibitemOpen
	\bibfield  {author} {\bibinfo {author} {\bibfnamefont {B.}~\bibnamefont
			{Chen}}, \bibinfo {author} {\bibfnamefont {Y.}~\bibnamefont {Wei}}, \bibinfo
		{author} {\bibfnamefont {T.}~\bibnamefont {Zhao}}, \bibinfo {author}
		{\bibfnamefont {S.}~\bibnamefont {Liu}}, \bibinfo {author} {\bibfnamefont
			{R.}~\bibnamefont {Su}}, \bibinfo {author} {\bibfnamefont {B.}~\bibnamefont
			{Yao}}, \bibinfo {author} {\bibfnamefont {Y.}~\bibnamefont {Yu}}, \bibinfo
		{author} {\bibfnamefont {J.}~\bibnamefont {Liu}}, \ and\ \bibinfo {author}
		{\bibfnamefont {X.}~\bibnamefont {Wang}},\ }\bibfield  {title} {\enquote
		{\bibinfo {title} {Bright solid-state sources for single photons with orbital
				angular momentum},}\ }\href {https://doi.org/10.1038/s41565-020-00827-7}
	{\bibfield  {journal} {\bibinfo  {journal} {Nature Nanotechnology}\ }\textbf
		{\bibinfo {volume} {16}},\ \bibinfo {pages} {302} (\bibinfo {year}
		{2021}{\natexlab{b}})}\BibitemShut {NoStop}%
	\bibitem [{\citenamefont {Liu}\ \emph {et~al.}(2024)\citenamefont {Liu},
		\citenamefont {Lao}, \citenamefont {Wang}, \citenamefont {Cheng},
		\citenamefont {Wang}, \citenamefont {Fu}, \citenamefont {Gao}, \citenamefont
		{Wang}, \citenamefont {Li}, \citenamefont {Gong} \emph
		{et~al.}}]{liu2024integrated}%
	\BibitemOpen
	\bibfield  {author} {\bibinfo {author} {\bibfnamefont {Y.}~\bibnamefont
			{Liu}}, \bibinfo {author} {\bibfnamefont {C.}~\bibnamefont {Lao}}, \bibinfo
		{author} {\bibfnamefont {M.}~\bibnamefont {Wang}}, \bibinfo {author}
		{\bibfnamefont {Y.}~\bibnamefont {Cheng}}, \bibinfo {author} {\bibfnamefont
			{Y.}~\bibnamefont {Wang}}, \bibinfo {author} {\bibfnamefont {S.}~\bibnamefont
			{Fu}}, \bibinfo {author} {\bibfnamefont {C.}~\bibnamefont {Gao}}, \bibinfo
		{author} {\bibfnamefont {J.}~\bibnamefont {Wang}}, \bibinfo {author}
		{\bibfnamefont {B.-B.}\ \bibnamefont {Li}}, \bibinfo {author} {\bibfnamefont
			{Q.}~\bibnamefont {Gong}},  \emph {et~al.},\ }\bibfield  {title} {\enquote
		{\bibinfo {title} {Integrated vortex soliton microcombs},}\ }\href
	{https://doi.org/10.1038/s41566-024-01418-x} {\bibfield  {journal} {\bibinfo
			{journal} {Nature Photonics}\ }\textbf {\bibinfo {volume} {18}},\ \bibinfo
		{pages} {632} (\bibinfo {year} {2024})}\BibitemShut {NoStop}%
	\bibitem [{\citenamefont {Chen}\ \emph {et~al.}(2024)\citenamefont {Chen},
		\citenamefont {Zhou}, \citenamefont {Liu}, \citenamefont {Ye}, \citenamefont
		{Cao}, \citenamefont {Huang}, \citenamefont {Kim}, \citenamefont {Zheng},
		\citenamefont {Oxenl{\o}we}, \citenamefont {Yvind} \emph
		{et~al.}}]{chen2024integrated}%
	\BibitemOpen
	\bibfield  {author} {\bibinfo {author} {\bibfnamefont {B.}~\bibnamefont
			{Chen}}, \bibinfo {author} {\bibfnamefont {Y.}~\bibnamefont {Zhou}}, \bibinfo
		{author} {\bibfnamefont {Y.}~\bibnamefont {Liu}}, \bibinfo {author}
		{\bibfnamefont {C.}~\bibnamefont {Ye}}, \bibinfo {author} {\bibfnamefont
			{Q.}~\bibnamefont {Cao}}, \bibinfo {author} {\bibfnamefont {P.}~\bibnamefont
			{Huang}}, \bibinfo {author} {\bibfnamefont {C.}~\bibnamefont {Kim}}, \bibinfo
		{author} {\bibfnamefont {Y.}~\bibnamefont {Zheng}}, \bibinfo {author}
		{\bibfnamefont {L.~K.}\ \bibnamefont {Oxenl{\o}we}}, \bibinfo {author}
		{\bibfnamefont {K.}~\bibnamefont {Yvind}},  \emph {et~al.},\ }\bibfield
	{title} {\enquote {\bibinfo {title} {Integrated optical vortex microcomb},}\
	}\href {https://doi.org/10.1038/s41566-024-01415-0} {\bibfield  {journal}
		{\bibinfo  {journal} {Nature Photonics}\ }\textbf {\bibinfo {volume} {18}},\
		\bibinfo {pages} {625} (\bibinfo {year} {2024})}\BibitemShut {NoStop}%
	\bibitem [{\citenamefont {Yang}\ \emph {et~al.}(2023)\citenamefont {Yang},
		\citenamefont {Xu}, \citenamefont {Wang}, \citenamefont {Zhang},
		\citenamefont {Li}, \citenamefont {Zhu}, \citenamefont {Wang}, \citenamefont
		{Dong}, \citenamefont {Fang}, \citenamefont {Yu}, \citenamefont {Guo},\ and\
		\citenamefont {Zou}}]{NOR2023}%
	\BibitemOpen
	\bibfield  {author} {\bibinfo {author} {\bibfnamefont {Y.-H.}\ \bibnamefont
			{Yang}}, \bibinfo {author} {\bibfnamefont {X.-B.}\ \bibnamefont {Xu}},
		\bibinfo {author} {\bibfnamefont {J.-Q.}\ \bibnamefont {Wang}}, \bibinfo
		{author} {\bibfnamefont {M.}~\bibnamefont {Zhang}}, \bibinfo {author}
		{\bibfnamefont {M.}~\bibnamefont {Li}}, \bibinfo {author} {\bibfnamefont
			{Z.-X.}\ \bibnamefont {Zhu}}, \bibinfo {author} {\bibfnamefont {Z.-B.}\
			\bibnamefont {Wang}}, \bibinfo {author} {\bibfnamefont {C.-H.}\ \bibnamefont
			{Dong}}, \bibinfo {author} {\bibfnamefont {W.}~\bibnamefont {Fang}}, \bibinfo
		{author} {\bibfnamefont {H.}~\bibnamefont {Yu}}, \bibinfo {author}
		{\bibfnamefont {G.-C.}\ \bibnamefont {Guo}}, \ and\ \bibinfo {author}
		{\bibfnamefont {C.-L.}\ \bibnamefont {Zou}},\ }\bibfield  {title} {\enquote
		{\bibinfo {title} {Nonlinear optical radiation of a lithium niobate
				microcavity},}\ }\href {\doibase 10.1103/PhysRevApplied.19.034087} {\bibfield
		{journal} {\bibinfo  {journal} {Physical Review Applied}\ }\textbf {\bibinfo
			{volume} {19}},\ \bibinfo {pages} {034087} (\bibinfo {year}
		{2023})}\BibitemShut {NoStop}%
	\bibitem [{\citenamefont {Li}\ \emph {et~al.}(2023)\citenamefont {Li},
		\citenamefont {Lin},\ and\ \citenamefont {Li}}]{li2023frequency}%
	\BibitemOpen
	\bibfield  {author} {\bibinfo {author} {\bibfnamefont {B.}~\bibnamefont
			{Li}}, \bibinfo {author} {\bibfnamefont {Q.}~\bibnamefont {Lin}}, \ and\
		\bibinfo {author} {\bibfnamefont {M.}~\bibnamefont {Li}},\ }\bibfield
	{title} {\enquote {\bibinfo {title} {Frequency--angular resolving lidar using
				chip-scale acousto-optic beam steering},}\ }\href
	{https://doi.org/10.1038/s41586-023-06201-6} {\bibfield  {journal} {\bibinfo
			{journal} {Nature}\ }\textbf {\bibinfo {volume} {620}},\ \bibinfo {pages}
		{316} (\bibinfo {year} {2023})}\BibitemShut {NoStop}%
	\bibitem [{\citenamefont {Liu}\ \emph {et~al.}(2022)\citenamefont {Liu},
		\citenamefont {Bo}, \citenamefont {Chang}, \citenamefont {Dong},
		\citenamefont {Ou}, \citenamefont {Regan}, \citenamefont {Shen},
		\citenamefont {Song}, \citenamefont {Yao}, \citenamefont {Zhang},
		\citenamefont {Zou},\ and\ \citenamefont {Xiao}}]{Liu2022}%
	\BibitemOpen
	\bibfield  {author} {\bibinfo {author} {\bibfnamefont {J.}~\bibnamefont
			{Liu}}, \bibinfo {author} {\bibfnamefont {F.}~\bibnamefont {Bo}}, \bibinfo
		{author} {\bibfnamefont {L.}~\bibnamefont {Chang}}, \bibinfo {author}
		{\bibfnamefont {C.-H.}\ \bibnamefont {Dong}}, \bibinfo {author}
		{\bibfnamefont {X.}~\bibnamefont {Ou}}, \bibinfo {author} {\bibfnamefont
			{B.}~\bibnamefont {Regan}}, \bibinfo {author} {\bibfnamefont
			{X.}~\bibnamefont {Shen}}, \bibinfo {author} {\bibfnamefont {Q.}~\bibnamefont
			{Song}}, \bibinfo {author} {\bibfnamefont {B.}~\bibnamefont {Yao}}, \bibinfo
		{author} {\bibfnamefont {W.}~\bibnamefont {Zhang}}, \bibinfo {author}
		{\bibfnamefont {C.-L.}\ \bibnamefont {Zou}}, \ and\ \bibinfo {author}
		{\bibfnamefont {Y.-F.}\ \bibnamefont {Xiao}},\ }\bibfield  {title} {\enquote
		{\bibinfo {title} {{Emerging material platforms for integrated microcavity
					photonics}},}\ }\href {\doibase 10.1007/s11433-022-1957-3} {\bibfield
		{journal} {\bibinfo  {journal} {Science China Physics, Mechanics \&
				Astronomy}\ }\textbf {\bibinfo {volume} {65}},\ \bibinfo {pages} {104201}
		(\bibinfo {year} {2022})}\BibitemShut {NoStop}%
	\bibitem [{\citenamefont {Xu}\ \emph {et~al.}(2025{\natexlab{a}})\citenamefont
		{Xu}, \citenamefont {Oudich}, \citenamefont {Zeng}, \citenamefont {Zhang},
		\citenamefont {Yang}, \citenamefont {Wang}, \citenamefont {Wang},
		\citenamefont {Sun}, \citenamefont {Guo}, \citenamefont {Jing},\ and\
		\citenamefont {Zou}}]{Xu2025a}%
	\BibitemOpen
	\bibfield  {author} {\bibinfo {author} {\bibfnamefont {X.-B.}\ \bibnamefont
			{Xu}}, \bibinfo {author} {\bibfnamefont {M.}~\bibnamefont {Oudich}}, \bibinfo
		{author} {\bibfnamefont {Y.}~\bibnamefont {Zeng}}, \bibinfo {author}
		{\bibfnamefont {J.-Z.}\ \bibnamefont {Zhang}}, \bibinfo {author}
		{\bibfnamefont {Y.-H.}\ \bibnamefont {Yang}}, \bibinfo {author}
		{\bibfnamefont {J.-Q.}\ \bibnamefont {Wang}}, \bibinfo {author}
		{\bibfnamefont {W.}~\bibnamefont {Wang}}, \bibinfo {author} {\bibfnamefont
			{L.}~\bibnamefont {Sun}}, \bibinfo {author} {\bibfnamefont {G.-C.}\
			\bibnamefont {Guo}}, \bibinfo {author} {\bibfnamefont {Y.}~\bibnamefont
			{Jing}}, \ and\ \bibinfo {author} {\bibfnamefont {C.-L.}\ \bibnamefont
			{Zou}},\ }\bibfield  {title} {\enquote {\bibinfo {title} {{Gigahertz
					topological phononic circuits based on micrometre-scale unsuspended waveguide
					arrays}},}\ }\href {\doibase 10.1038/s41928-025-01437-8} {\bibfield
		{journal} {\bibinfo  {journal} {Nature Electronics}\ }\textbf {\bibinfo
			{volume} {8}},\ \bibinfo {pages} {689} (\bibinfo {year}
		{2025}{\natexlab{a}})}\BibitemShut {NoStop}%
	\bibitem [{\citenamefont {Yang}\ \emph
		{et~al.}(2024{\natexlab{a}})\citenamefont {Yang}, \citenamefont {Wang},
		\citenamefont {Xu}, \citenamefont {Li}, \citenamefont {Zhang}, \citenamefont
		{Pan}, \citenamefont {Xiao}, \citenamefont {Wang}, \citenamefont {Guo},
		\citenamefont {Sun},\ and\ \citenamefont {Zou}}]{Yang2024ome}%
	\BibitemOpen
	\bibfield  {author} {\bibinfo {author} {\bibfnamefont {Y.-H.}\ \bibnamefont
			{Yang}}, \bibinfo {author} {\bibfnamefont {J.-Q.}\ \bibnamefont {Wang}},
		\bibinfo {author} {\bibfnamefont {X.-B.}\ \bibnamefont {Xu}}, \bibinfo
		{author} {\bibfnamefont {M.}~\bibnamefont {Li}}, \bibinfo {author}
		{\bibfnamefont {Y.-L.}\ \bibnamefont {Zhang}}, \bibinfo {author}
		{\bibfnamefont {X.}~\bibnamefont {Pan}}, \bibinfo {author} {\bibfnamefont
			{L.}~\bibnamefont {Xiao}}, \bibinfo {author} {\bibfnamefont {W.}~\bibnamefont
			{Wang}}, \bibinfo {author} {\bibfnamefont {G.-C.}\ \bibnamefont {Guo}},
		\bibinfo {author} {\bibfnamefont {L.}~\bibnamefont {Sun}}, \ and\ \bibinfo
		{author} {\bibfnamefont {C.-l.}\ \bibnamefont {Zou}},\ }\bibfield  {title}
	{\enquote {\bibinfo {title} {{Proposal for Brillouin microwave-to-optical
					conversion on a chip [Invited]}},}\ }\href {\doibase 10.1364/OME.534817}
	{\bibfield  {journal} {\bibinfo  {journal} {Optical Materials Express}\
		}\textbf {\bibinfo {volume} {14}},\ \bibinfo {pages} {2400} (\bibinfo {year}
		{2024}{\natexlab{a}})}\BibitemShut {NoStop}%
	\bibitem [{\citenamefont {Ye}\ \emph {et~al.}(2025)\citenamefont {Ye},
		\citenamefont {Feng}, \citenamefont {Te~Morsche}, \citenamefont {Wei},
		\citenamefont {Klaver}, \citenamefont {Mishra}, \citenamefont {Zheng},
		\citenamefont {Keloth}, \citenamefont {Tarik~Isik}, \citenamefont {Chen},\
		and\ \citenamefont {Others}}]{Ye2025}%
	\BibitemOpen
	\bibfield  {author} {\bibinfo {author} {\bibfnamefont {K.}~\bibnamefont
			{Ye}}, \bibinfo {author} {\bibfnamefont {H.}~\bibnamefont {Feng}}, \bibinfo
		{author} {\bibfnamefont {R.}~\bibnamefont {Te~Morsche}}, \bibinfo {author}
		{\bibfnamefont {C.}~\bibnamefont {Wei}}, \bibinfo {author} {\bibfnamefont
			{Y.}~\bibnamefont {Klaver}}, \bibinfo {author} {\bibfnamefont
			{A.}~\bibnamefont {Mishra}}, \bibinfo {author} {\bibfnamefont
			{Z.}~\bibnamefont {Zheng}}, \bibinfo {author} {\bibfnamefont
			{A.}~\bibnamefont {Keloth}}, \bibinfo {author} {\bibfnamefont
			{A.}~\bibnamefont {Tarik~Isik}}, \bibinfo {author} {\bibfnamefont
			{Z.}~\bibnamefont {Chen}}, \ and\ \bibinfo {author} {\bibnamefont {Others}},\
	}\bibfield  {title} {\enquote {\bibinfo {title} {{Integrated Brillouin
					photonics in thin-film lithium niobate}},}\ }\href
	{https://www.science.org/doi/10.1126/sciadv.adv4022} {\bibfield  {journal}
		{\bibinfo  {journal} {Science Advances}\ }\textbf {\bibinfo {volume} {11}},\
		\bibinfo {pages} {eadv4022} (\bibinfo {year} {2025})}\BibitemShut {NoStop}%
	\bibitem [{\citenamefont {Rodrigues}\ \emph {et~al.}(2025)\citenamefont
		{Rodrigues}, \citenamefont {Schilder}, \citenamefont {Zurita}, \citenamefont
		{Magalh{\~{a}}es}, \citenamefont {Shams-Ansari}, \citenamefont {dos Santos},
		\citenamefont {Paiano}, \citenamefont {Alegre}, \citenamefont
		{Lon{\v{c}}ar},\ and\ \citenamefont {Wiederhecker}}]{Rodrigues2025}%
	\BibitemOpen
	\bibfield  {author} {\bibinfo {author} {\bibfnamefont {C.~C.}\ \bibnamefont
			{Rodrigues}}, \bibinfo {author} {\bibfnamefont {N.~J.}\ \bibnamefont
			{Schilder}}, \bibinfo {author} {\bibfnamefont {R.~O.}\ \bibnamefont
			{Zurita}}, \bibinfo {author} {\bibfnamefont {L.~S.}\ \bibnamefont
			{Magalh{\~{a}}es}}, \bibinfo {author} {\bibfnamefont {A.}~\bibnamefont
			{Shams-Ansari}}, \bibinfo {author} {\bibfnamefont {F.~J.~L.}\ \bibnamefont
			{dos Santos}}, \bibinfo {author} {\bibfnamefont {O.~M.}\ \bibnamefont
			{Paiano}}, \bibinfo {author} {\bibfnamefont {T.~P.~M.}\ \bibnamefont
			{Alegre}}, \bibinfo {author} {\bibfnamefont {M.}~\bibnamefont
			{Lon{\v{c}}ar}}, \ and\ \bibinfo {author} {\bibfnamefont {G.~S.}\
			\bibnamefont {Wiederhecker}},\ }\bibfield  {title} {\enquote {\bibinfo
			{title} {{Cross-Polarized Stimulated Brillouin Scattering in Lithium Niobate
					Waveguides}},}\ }\href {\doibase 10.1103/PhysRevLett.134.113601} {\bibfield
		{journal} {\bibinfo  {journal} {Physical Review Letters}\ }\textbf {\bibinfo
			{volume} {134}},\ \bibinfo {pages} {113601} (\bibinfo {year}
		{2025})}\BibitemShut {NoStop}%
	\bibitem [{\citenamefont {Yu}\ \emph {et~al.}(2025)\citenamefont {Yu},
		\citenamefont {Zhou}, \citenamefont {Yang}, \citenamefont {Zhang},
		\citenamefont {Zhu}, \citenamefont {Yang}, \citenamefont {Xu}, \citenamefont
		{Chen}, \citenamefont {Zou},\ and\ \citenamefont {Lu}}]{Yu2025}%
	\BibitemOpen
	\bibfield  {author} {\bibinfo {author} {\bibfnamefont {S.}~\bibnamefont
			{Yu}}, \bibinfo {author} {\bibfnamefont {R.}~\bibnamefont {Zhou}}, \bibinfo
		{author} {\bibfnamefont {G.}~\bibnamefont {Yang}}, \bibinfo {author}
		{\bibfnamefont {Q.}~\bibnamefont {Zhang}}, \bibinfo {author} {\bibfnamefont
			{H.}~\bibnamefont {Zhu}}, \bibinfo {author} {\bibfnamefont {Y.}~\bibnamefont
			{Yang}}, \bibinfo {author} {\bibfnamefont {X.-B.}\ \bibnamefont {Xu}},
		\bibinfo {author} {\bibfnamefont {B.}~\bibnamefont {Chen}}, \bibinfo {author}
		{\bibfnamefont {C.-L.}\ \bibnamefont {Zou}}, \ and\ \bibinfo {author}
		{\bibfnamefont {J.}~\bibnamefont {Lu}},\ }\bibfield  {title} {\enquote
		{\bibinfo {title} {On-chip {Brillouin} amplifier in suspended lithium niobate
				nanowaveguides},}\ }\href
	{https://onlinelibrary.wiley.com/doi/10.1002/lpor.202500027} {\bibfield
		{journal} {\bibinfo  {journal} {Laser \& Photonics Reviews}\ ,\ \bibinfo
			{pages} {2500027}} (\bibinfo {year} {2025})}\BibitemShut {NoStop}%
	\bibitem [{\citenamefont {Yang}\ \emph
		{et~al.}(2025{\natexlab{a}})\citenamefont {Yang}, \citenamefont {Wang},
		\citenamefont {Zhu}, \citenamefont {Zeng}, \citenamefont {Li}, \citenamefont
		{Zhang}, \citenamefont {Lu}, \citenamefont {Zhang}, \citenamefont {Wang},
		\citenamefont {Dong}, \citenamefont {Xu}, \citenamefont {Guo}, \citenamefont
		{Sun},\ and\ \citenamefont {Zou}}]{Yang2025a}%
	\BibitemOpen
	\bibfield  {author} {\bibinfo {author} {\bibfnamefont {Y.-H.}\ \bibnamefont
			{Yang}}, \bibinfo {author} {\bibfnamefont {J.-Q.}\ \bibnamefont {Wang}},
		\bibinfo {author} {\bibfnamefont {Z.-X.}\ \bibnamefont {Zhu}}, \bibinfo
		{author} {\bibfnamefont {Y.}~\bibnamefont {Zeng}}, \bibinfo {author}
		{\bibfnamefont {M.}~\bibnamefont {Li}}, \bibinfo {author} {\bibfnamefont
			{Y.-L.}\ \bibnamefont {Zhang}}, \bibinfo {author} {\bibfnamefont
			{J.}~\bibnamefont {Lu}}, \bibinfo {author} {\bibfnamefont {Q.}~\bibnamefont
			{Zhang}}, \bibinfo {author} {\bibfnamefont {W.}~\bibnamefont {Wang}},
		\bibinfo {author} {\bibfnamefont {C.-H.}\ \bibnamefont {Dong}}, \bibinfo
		{author} {\bibfnamefont {X.-B.}\ \bibnamefont {Xu}}, \bibinfo {author}
		{\bibfnamefont {G.-C.}\ \bibnamefont {Guo}}, \bibinfo {author} {\bibfnamefont
			{L.}~\bibnamefont {Sun}}, \ and\ \bibinfo {author} {\bibfnamefont {C.-L.}\
			\bibnamefont {Zou}},\ }\bibfield  {title} {\enquote {\bibinfo {title}
			{{Multi-Channel Microwave-to-Optics Conversion Utilizing a Hybrid
					Photonic-Phononic Waveguide}},}\ }\href {http://arxiv.org/abs/2509.10052}
	{\bibfield  {journal} {\bibinfo  {journal} {arXiv preprint: 2509.10052}\ }
		(\bibinfo {year} {2025}{\natexlab{a}})}\BibitemShut {NoStop}%
	\bibitem [{\citenamefont {Xu}\ \emph {et~al.}(2025{\natexlab{b}})\citenamefont
		{Xu}, \citenamefont {Zhu}, \citenamefont {Yang}, \citenamefont {Wang},
		\citenamefont {Zeng}, \citenamefont {Zou}, \citenamefont {Lu}, \citenamefont
		{Zhang}, \citenamefont {Wang}, \citenamefont {Guo}, \citenamefont {Sun},\
		and\ \citenamefont {Zou}}]{Xu2025b}%
	\BibitemOpen
	\bibfield  {author} {\bibinfo {author} {\bibfnamefont {X.-B.}\ \bibnamefont
			{Xu}}, \bibinfo {author} {\bibfnamefont {Z.-X.}\ \bibnamefont {Zhu}},
		\bibinfo {author} {\bibfnamefont {Y.-H.}\ \bibnamefont {Yang}}, \bibinfo
		{author} {\bibfnamefont {J.-Q.}\ \bibnamefont {Wang}}, \bibinfo {author}
		{\bibfnamefont {Y.}~\bibnamefont {Zeng}}, \bibinfo {author} {\bibfnamefont
			{J.-H.}\ \bibnamefont {Zou}}, \bibinfo {author} {\bibfnamefont
			{J.}~\bibnamefont {Lu}}, \bibinfo {author} {\bibfnamefont {Y.-L.}\
			\bibnamefont {Zhang}}, \bibinfo {author} {\bibfnamefont {W.}~\bibnamefont
			{Wang}}, \bibinfo {author} {\bibfnamefont {G.-C.}\ \bibnamefont {Guo}},
		\bibinfo {author} {\bibfnamefont {L.}~\bibnamefont {Sun}}, \ and\ \bibinfo
		{author} {\bibfnamefont {C.-L.}\ \bibnamefont {Zou}},\ }\bibfield  {title}
	{\enquote {\bibinfo {title} {{Magnetic-free optical mode degeneracy lifting
					in lithium niobate microring resonators}},}\ }\href
	{http://arxiv.org/abs/2509.01940} {\bibfield  {journal} {\bibinfo  {journal}
			{arXiv preprint: 2509.01940}\ } (\bibinfo {year}
		{2025}{\natexlab{b}})}\BibitemShut {NoStop}%
	\bibitem [{\citenamefont {Chen}\ \emph
		{et~al.}(2025{\natexlab{a}})\citenamefont {Chen}, \citenamefont {Yang},
		\citenamefont {Zhu}, \citenamefont {Wang}, \citenamefont {Xu}, \citenamefont
		{Li}, \citenamefont {Han}, \citenamefont {Guo}, \citenamefont {Chen},\ and\
		\citenamefont {Zou}}]{Chen2025}%
	\BibitemOpen
	\bibfield  {author} {\bibinfo {author} {\bibfnamefont {J.-L.}\ \bibnamefont
			{Chen}}, \bibinfo {author} {\bibfnamefont {Y.-H.}\ \bibnamefont {Yang}},
		\bibinfo {author} {\bibfnamefont {Z.-X.}\ \bibnamefont {Zhu}}, \bibinfo
		{author} {\bibfnamefont {J.-Q.}\ \bibnamefont {Wang}}, \bibinfo {author}
		{\bibfnamefont {X.-B.}\ \bibnamefont {Xu}}, \bibinfo {author} {\bibfnamefont
			{M.}~\bibnamefont {Li}}, \bibinfo {author} {\bibfnamefont {Z.-F.}\
			\bibnamefont {Han}}, \bibinfo {author} {\bibfnamefont {G.-C.}\ \bibnamefont
			{Guo}}, \bibinfo {author} {\bibfnamefont {W.}~\bibnamefont {Chen}}, \ and\
		\bibinfo {author} {\bibfnamefont {C.-L.}\ \bibnamefont {Zou}},\ }\bibfield
	{title} {\enquote {\bibinfo {title} {{Proposal for Forward Brillouin
					Inter-Modal Scattering in Non-suspended Lithium Niobate Waveguides at Visible
					Wavelengths}},}\ }\href {http://arxiv.org/abs/2510.08170} {\bibfield
		{journal} {\bibinfo  {journal} {arXiv preprint: 2510.08170}\ } (\bibinfo
		{year} {2025}{\natexlab{a}})}\BibitemShut {NoStop}%
	\bibitem [{\citenamefont {Yang}\ \emph
		{et~al.}(2024{\natexlab{b}})\citenamefont {Yang}, \citenamefont {Wang},
		\citenamefont {Zhu}, \citenamefont {Xu}, \citenamefont {Zhang}, \citenamefont
		{Lu}, \citenamefont {Zeng}, \citenamefont {Dong}, \citenamefont {Sun},
		\citenamefont {Guo},\ and\ \citenamefont {Zou}}]{Yang2023}%
	\BibitemOpen
	\bibfield  {author} {\bibinfo {author} {\bibfnamefont {Y.-H.}\ \bibnamefont
			{Yang}}, \bibinfo {author} {\bibfnamefont {J.-Q.}\ \bibnamefont {Wang}},
		\bibinfo {author} {\bibfnamefont {Z.-X.}\ \bibnamefont {Zhu}}, \bibinfo
		{author} {\bibfnamefont {X.-B.}\ \bibnamefont {Xu}}, \bibinfo {author}
		{\bibfnamefont {Q.}~\bibnamefont {Zhang}}, \bibinfo {author} {\bibfnamefont
			{J.}~\bibnamefont {Lu}}, \bibinfo {author} {\bibfnamefont {Y.}~\bibnamefont
			{Zeng}}, \bibinfo {author} {\bibfnamefont {C.-H.}\ \bibnamefont {Dong}},
		\bibinfo {author} {\bibfnamefont {L.}~\bibnamefont {Sun}}, \bibinfo {author}
		{\bibfnamefont {G.-C.}\ \bibnamefont {Guo}}, \ and\ \bibinfo {author}
		{\bibfnamefont {C.-L.}\ \bibnamefont {Zou}},\ }\bibfield  {title} {\enquote
		{\bibinfo {title} {{Stimulated Brillouin interaction between guided phonons
					and photons in a lithium niobate waveguide}},}\ }\href {\doibase
		10.1007/s11433-023-2272-y} {\bibfield  {journal} {\bibinfo  {journal}
			{Science China Physics, Mechanics \& Astronomy}\ }\textbf {\bibinfo {volume}
			{67}},\ \bibinfo {pages} {214221} (\bibinfo {year}
		{2024}{\natexlab{b}})}\BibitemShut {NoStop}%
	\bibitem [{sm()}]{sm}%
	\BibitemOpen
	\bibfield  {title} {\enquote {\bibinfo {title} {See the supplemental
				materials for details about theoretical derivations of {Brillouin} nonlinear
				optical radiation.}}\ }\href@noop {} {\ }\BibitemShut {NoStop}%
	\bibitem [{\citenamefont {Wang}\ \emph {et~al.}(2024)\citenamefont {Wang},
		\citenamefont {Hu}, \citenamefont {Lao}, \citenamefont {Wang}, \citenamefont
		{Jin}, \citenamefont {Zhou}, \citenamefont {Lei}, \citenamefont {Wang},
		\citenamefont {Liu}, \citenamefont {Yang},\ and\ \citenamefont
		{Li}}]{PhysRevX.14.011056}%
	\BibitemOpen
	\bibfield  {author} {\bibinfo {author} {\bibfnamefont {M.}~\bibnamefont
			{Wang}}, \bibinfo {author} {\bibfnamefont {Z.-G.}\ \bibnamefont {Hu}},
		\bibinfo {author} {\bibfnamefont {C.}~\bibnamefont {Lao}}, \bibinfo {author}
		{\bibfnamefont {Y.}~\bibnamefont {Wang}}, \bibinfo {author} {\bibfnamefont
			{X.}~\bibnamefont {Jin}}, \bibinfo {author} {\bibfnamefont {X.}~\bibnamefont
			{Zhou}}, \bibinfo {author} {\bibfnamefont {Y.}~\bibnamefont {Lei}}, \bibinfo
		{author} {\bibfnamefont {Z.}~\bibnamefont {Wang}}, \bibinfo {author}
		{\bibfnamefont {W.}~\bibnamefont {Liu}}, \bibinfo {author} {\bibfnamefont
			{Q.-F.}\ \bibnamefont {Yang}}, \ and\ \bibinfo {author} {\bibfnamefont
			{B.-B.}\ \bibnamefont {Li}},\ }\bibfield  {title} {\enquote {\bibinfo {title}
			{Taming {Brillouin} optomechanics using supermode microresonators},}\ }\href
	{\doibase 10.1103/PhysRevX.14.011056} {\bibfield  {journal} {\bibinfo
			{journal} {Physical Review X}\ }\textbf {\bibinfo {volume} {14}},\ \bibinfo
		{pages} {011056} (\bibinfo {year} {2024})}\BibitemShut {NoStop}%
	\bibitem [{\citenamefont {Yang}\ \emph
		{et~al.}(2025{\natexlab{b}})\citenamefont {Yang}, \citenamefont {Wang},
		\citenamefont {Zhu}, \citenamefont {Xu}, \citenamefont {Li}, \citenamefont
		{Lu}, \citenamefont {Guo}, \citenamefont {Sun},\ and\ \citenamefont
		{Zou}}]{Yang2025b}%
	\BibitemOpen
	\bibfield  {author} {\bibinfo {author} {\bibfnamefont {Y.-H.}\ \bibnamefont
			{Yang}}, \bibinfo {author} {\bibfnamefont {J.-Q.}\ \bibnamefont {Wang}},
		\bibinfo {author} {\bibfnamefont {Z.-X.}\ \bibnamefont {Zhu}}, \bibinfo
		{author} {\bibfnamefont {X.-B.}\ \bibnamefont {Xu}}, \bibinfo {author}
		{\bibfnamefont {M.}~\bibnamefont {Li}}, \bibinfo {author} {\bibfnamefont
			{J.}~\bibnamefont {Lu}}, \bibinfo {author} {\bibfnamefont {G.-C.}\
			\bibnamefont {Guo}}, \bibinfo {author} {\bibfnamefont {L.}~\bibnamefont
			{Sun}}, \ and\ \bibinfo {author} {\bibfnamefont {C.-L.}\ \bibnamefont
			{Zou}},\ }\bibfield  {title} {\enquote {\bibinfo {title} {{Suspension-Free
					Integrated Cavity Brillouin Optomechanics on a Chip}},}\ }\href
	{http://arxiv.org/abs/2510.20463} {\bibfield  {journal} {\bibinfo  {journal}
			{arXiv preprint: 2510.20463}\ } (\bibinfo {year}
		{2025}{\natexlab{b}})}\BibitemShut {NoStop}%
	\bibitem [{\citenamefont {Chen}\ \emph
		{et~al.}(2025{\natexlab{b}})\citenamefont {Chen}, \citenamefont {Yama},
		\citenamefont {Deng}, \citenamefont {Lin}, \citenamefont {Yu}, \citenamefont
		{Saxena}, \citenamefont {Majumdar}, \citenamefont {Fu},\ and\ \citenamefont
		{Li}}]{Chen2025silicon}%
	\BibitemOpen
	\bibfield  {author} {\bibinfo {author} {\bibfnamefont {I.-T.}\ \bibnamefont
			{Chen}}, \bibinfo {author} {\bibfnamefont {N.~S.}\ \bibnamefont {Yama}},
		\bibinfo {author} {\bibfnamefont {H.}~\bibnamefont {Deng}}, \bibinfo {author}
		{\bibfnamefont {Q.}~\bibnamefont {Lin}}, \bibinfo {author} {\bibfnamefont
			{Y.}~\bibnamefont {Yu}}, \bibinfo {author} {\bibfnamefont {A.}~\bibnamefont
			{Saxena}}, \bibinfo {author} {\bibfnamefont {A.}~\bibnamefont {Majumdar}},
		\bibinfo {author} {\bibfnamefont {K.-M.~C.}\ \bibnamefont {Fu}}, \ and\
		\bibinfo {author} {\bibfnamefont {M.}~\bibnamefont {Li}},\ }\bibfield
	{title} {\enquote {\bibinfo {title} {{Intermodal microwave-to-optical
					transduction using silicon-on-sapphire optomechanical ring resonator}},}\
	}\href {\doibase 10.1126/sciadv.adx6485} {\bibfield  {journal} {\bibinfo
			{journal} {Science Advances}\ }\textbf {\bibinfo {volume} {11}},\ \bibinfo
		{pages} {eadx6485} (\bibinfo {year} {2025}{\natexlab{b}})}\BibitemShut
	{NoStop}%
	\bibitem [{\citenamefont {Fu}\ \emph {et~al.}(2019)\citenamefont {Fu},
		\citenamefont {Shen}, \citenamefont {Xu}, \citenamefont {Zou}, \citenamefont
		{Cheng}, \citenamefont {Han},\ and\ \citenamefont {Tang}}]{Fu2019}%
	\BibitemOpen
	\bibfield  {author} {\bibinfo {author} {\bibfnamefont {W.}~\bibnamefont
			{Fu}}, \bibinfo {author} {\bibfnamefont {Z.}~\bibnamefont {Shen}}, \bibinfo
		{author} {\bibfnamefont {Y.}~\bibnamefont {Xu}}, \bibinfo {author}
		{\bibfnamefont {C.-L.}\ \bibnamefont {Zou}}, \bibinfo {author} {\bibfnamefont
			{R.}~\bibnamefont {Cheng}}, \bibinfo {author} {\bibfnamefont
			{X.}~\bibnamefont {Han}}, \ and\ \bibinfo {author} {\bibfnamefont {H.~X.}\
			\bibnamefont {Tang}},\ }\bibfield  {title} {\enquote {\bibinfo {title}
			{{Phononic integrated circuitry and spin–orbit interaction of phonons}},}\
	}\href {\doibase 10.1038/s41467-019-10852-3} {\bibfield  {journal} {\bibinfo
			{journal} {Nature Communications}\ }\textbf {\bibinfo {volume} {10}},\
		\bibinfo {pages} {2743} (\bibinfo {year} {2019})}\BibitemShut {NoStop}%
	\bibitem [{\citenamefont {Zhang}\ \emph {et~al.}(2025)\citenamefont {Zhang},
		\citenamefont {Cui}, \citenamefont {Xue}, \citenamefont {Chen},\ and\
		\citenamefont {Fan}}]{Zhang2025}%
	\BibitemOpen
	\bibfield  {author} {\bibinfo {author} {\bibfnamefont {L.}~\bibnamefont
			{Zhang}}, \bibinfo {author} {\bibfnamefont {C.}~\bibnamefont {Cui}}, \bibinfo
		{author} {\bibfnamefont {Y.}~\bibnamefont {Xue}}, \bibinfo {author}
		{\bibfnamefont {P.}~\bibnamefont {Chen}}, \ and\ \bibinfo {author}
		{\bibfnamefont {L.}~\bibnamefont {Fan}},\ }\bibfield  {title} {\enquote
		{\bibinfo {title} {{Scalable photonic-phonoinc integrated circuitry for
					reconfigurable signal processing}},}\ }\href {\doibase
		10.1038/s41467-025-57922-3} {\bibfield  {journal} {\bibinfo  {journal}
			{Nature Communications}\ }\textbf {\bibinfo {volume} {16}},\ \bibinfo {pages}
		{2718} (\bibinfo {year} {2025})}\BibitemShut {NoStop}%
	\bibitem [{\citenamefont {Xu}\ \emph {et~al.}(2022)\citenamefont {Xu},
		\citenamefont {Wang}, \citenamefont {Yang}, \citenamefont {Wang},
		\citenamefont {Zhang}, \citenamefont {Wang}, \citenamefont {Dong},
		\citenamefont {Sun}, \citenamefont {Guo},\ and\ \citenamefont
		{Zou}}]{Xu2022}%
	\BibitemOpen
	\bibfield  {author} {\bibinfo {author} {\bibfnamefont {X.-B.}\ \bibnamefont
			{Xu}}, \bibinfo {author} {\bibfnamefont {J.-Q.}\ \bibnamefont {Wang}},
		\bibinfo {author} {\bibfnamefont {Y.-H.}\ \bibnamefont {Yang}}, \bibinfo
		{author} {\bibfnamefont {W.}~\bibnamefont {Wang}}, \bibinfo {author}
		{\bibfnamefont {Y.-L.}\ \bibnamefont {Zhang}}, \bibinfo {author}
		{\bibfnamefont {B.-Z.}\ \bibnamefont {Wang}}, \bibinfo {author}
		{\bibfnamefont {C.-H.}\ \bibnamefont {Dong}}, \bibinfo {author}
		{\bibfnamefont {L.}~\bibnamefont {Sun}}, \bibinfo {author} {\bibfnamefont
			{G.-C.}\ \bibnamefont {Guo}}, \ and\ \bibinfo {author} {\bibfnamefont
			{C.-L.}\ \bibnamefont {Zou}},\ }\bibfield  {title} {\enquote {\bibinfo
			{title} {{High-frequency traveling-wave phononic cavity with sub-micron
					wavelength}},}\ }\href {\doibase 10.1063/5.0086751} {\bibfield  {journal}
		{\bibinfo  {journal} {Applied Physics Letters}\ }\textbf {\bibinfo {volume}
			{120}},\ \bibinfo {pages} {163503} (\bibinfo {year} {2022})}\BibitemShut
	{NoStop}%
\end{thebibliography}
\end{document}